\newcommand{\schemename}{SmartGuard}
\documentclass[sigconf]{acmart}
\AtBeginDocument{%
  \providecommand\BibTeX{{%
    \normalfont B\kern-0.5em{\scshape i\kern-0.25em b}\kern-0.8em\TeX}}}

\copyrightyear{2024}
\acmYear{2024}
\setcopyright{acmlicensed}\acmConference[KDD '24]{Proceedings of the 30th
ACM SIGKDD Conference on Knowledge Discovery and Data Mining}{August
25--29, 2024}{Barcelona, Spain}
\acmBooktitle{Proceedings of the 30th ACM SIGKDD Conference on Knowledge
Discovery and Data Mining (KDD '24), August 25--29, 2024, Barcelona, Spain}
\acmDOI{10.1145/3637528.3671708}
\acmISBN{979-8-4007-0490-1/24/08}

\makeatletter
\gdef\@copyrightpermission{
  \begin{minipage}{0.3\columnwidth}
   \href{https://creativecommons.org/licenses/by/4.0/}{\includegraphics[width=0.90\textwidth]{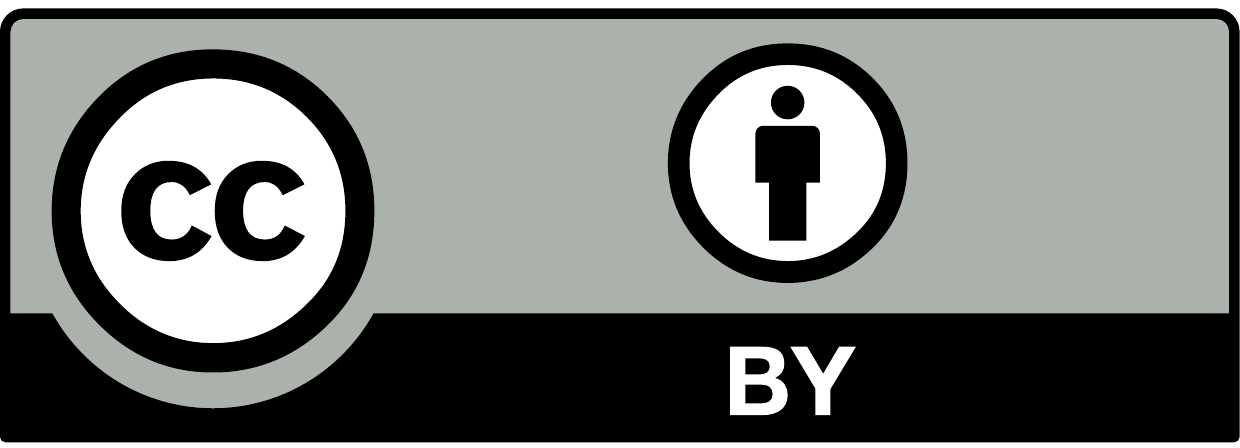}}
  \end{minipage}\hfill
  \begin{minipage}{0.7\columnwidth}
   \href{https://creativecommons.org/licenses/by/4.0/}{This work is licensed under a Creative Commons Attribution International 4.0 License.}
  \end{minipage}
  \vspace{5pt}
}
\makeatother

\usepackage{algorithmic}
\usepackage[ruled, linesnumbered]{algorithm2e}
\usepackage{subfigure}
\usepackage{multirow}

\newtheorem{myDef}{Definition}
\newtheorem{myPro}{Problem}
\newcommand{\problem}{UBS}

\begin{document}


\title{Make Your Home Safe: Time-aware Unsupervised User Behavior Anomaly Detection in Smart Homes via Loss-guided Mask}



\author{Jingyu Xiao}
\affiliation{
  \institution{Peng Cheng Laboratory}
  \institution{Tsinghua Shenzhen International Graduate School}
  \city{Shenzhen}
  \country{China}}
\email{jy-xiao21@mails.tsinghua.edu.cn
}

\authornotemark[1]
\author{Zhiyao Xu}
\affiliation{
  \institution{Xi'an University of Electronic Science and Technology}
  \city{Xi'an}
  \country{China}}
\email{21009200843@stu.xidian.edu.cn}

\authornotemark[1]
\author{Qingsong Zou}
\affiliation{
  \institution{Tsinghua Shenzhen International Graduate School}
  \institution{Peng Cheng Laboratory}
  \city{Shenzhen}
  \country{China}}
\email{zouqs21@mails.tsinghua.edu.cn
}
\authornote{The first three authors have equal contribution.}

\author{Qing Li}
\affiliation{
  \institution{Peng Cheng Laboratory}
  \city{Shenzhen}
  \country{China}}
\email{liq@pcl.ac.cn}
\authornote{Qing Li is the corresponding author.}

\author{Dan Zhao}
\affiliation{
  \institution{Peng Cheng Laboratory}
  \city{Shenzhen}
  \country{China}}
\email{zhaod01@pcl.ac.cn}

\author{Dong Fang}
\affiliation{
  \institution{Tencent}
  \city{Shenzhen}
  \country{China}}
\email{victordfang@tencent.com}

\author{Ruoyu Li}
\affiliation{
  \institution{Tsinghua Shenzhen International Graduate School}
  \city{Shenzhen}
  \country{China}}
\email{liry19@mails.tsinghua.edu.cn}

\author{Wenxin Tang}
\affiliation{
  \institution{Tsinghua Shenzhen International Graduate School}
  \city{Shenzhen}
  \country{China}}
\email{vinsontang2126@gmail.com}

\author{Kang Li}
\affiliation{
  \institution{Tsinghua Shenzhen International Graduate School}
  \city{Shenzhen}
  \country{China}}
\email{lk26603878@gmail.com}

\author{Xudong Zuo}
\affiliation{
  \institution{Tsinghua Shenzhen International Graduate School}
  \city{Shenzhen}
  \country{China}}
\email{zuoxd20@mails.tsinghua.edu.cn}

\author{Penghui Hu}
\affiliation{
  \institution{Tsinghua University}
  \city{Beijing}
  \country{China}}
\email{huph22@mails.tsinghua.edu.cn}

\author{Yong Jiang}
\affiliation{
  \institution{Tsinghua Shenzhen International Graduate School}
  \institution{Peng Cheng Laboratory}
  \city{Shenzhen}
  \country{China}}
\email{jiangy@sz.tsinghua.edu.cn}

\author{Zixuan Weng}
\affiliation{
  \institution{Beijing Jiaotong University}
  \city{Beijing}
  \country{China}}
\email{20722027@bjtu.edu.cn}

\author{Michael R.Lyu}
\affiliation{
  \institution{The Chinese University of Hong Kong}
  \city{Hong Kong}
  \country{China}}
\email{lyu@cse.cuhk.edu.hk}




\renewcommand{\shortauthors}{Jingyu Xiao, Zhiyao Xu and Qingsong Zou et al.}

\begin{abstract}
Smart homes, powered by the Internet of Things, offer great convenience but also pose security concerns due to abnormal behaviors, such as improper operations of users and potential attacks from malicious attackers. Several behavior modeling methods have been proposed to identify abnormal behaviors and mitigate potential risks. However, their performance often falls short because they do not effectively learn less frequent behaviors, consider temporal context, or account for the impact of noise in human behaviors. In this paper, we propose \schemename, an autoencoder-based unsupervised user behavior anomaly detection framework. First, we design a Loss-guided Dynamic Mask Strategy (LDMS) to encourage the model to learn less frequent behaviors, which are often overlooked during learning. Second, we propose a Three-level Time-aware Position Embedding (TTPE) to incorporate temporal information into positional embedding to detect temporal context anomaly. Third, we propose a Noise-aware Weighted Reconstruction Loss (NWRL) that assigns different weights for routine behaviors and noise behaviors to mitigate the interference of noise behaviors during inference. 
Comprehensive experiments demonstrate that \schemename \ consistently outperforms state-of-the-art baselines and also offers highly interpretable results.

\end{abstract}






\maketitle

\section{Introduction}
\label{sec:intro}



The rapid growth of IoT solutions has led to an unprecedented increase in smart devices within homes, expected to reach approximately 5 billion by 2025 \cite{iot-analytics}. However, the abnormal behaviors pose substantial security risks within smart homes. These abnormal behaviors usually originate from two primary sources. First, improper operations by users can cause abnormal behaviors, such as inadvertently activating the air conditioner's cooling mode during winter or forgetting to close a water valve. Second, malicious attackers can exploit vulnerabilities within IoT devices and platforms, taking unauthorized control of these devices. For example, hackers can compromise IoT platforms, allowing them to disable security cameras and manipulate home automation systems, creating opportunities for burglary. These security concerns emphasize the urgency of robust behavioral modeling methods and enhanced security measures to safeguard smart home environments.


Deep learning has been employed across various domains to mine correlations between behaviors for modeling user behavior sequences \cite{shisong2022, shisong2023, shisong2024} and address security issues \cite{gao2023backdoor, gao2024energy, gao2024inducing, bai2023badclip}. DeepMove\cite{feng2018deepmove} leverages RNNs to model both long and short-term mobility patterns of users for human mobility prediction. To capture the dynamics of user's behaviors, SASRec \cite{kang2018self} proposes a self-attention based model to achieve sequential recommendation. More recent efforts\cite{chen2019behavior, sun2019bert4rec, de2021transformers4rec} primarily focus on transformer-based models for their superior ability to handle sequential behavior data.



However, we cannot borrow the above models to directly apply them in our scenarios, because of the following three challenges of user behavior modeling in smart homes.



\begin{figure}[ht]
\centering
\includegraphics[width = .45\textwidth]{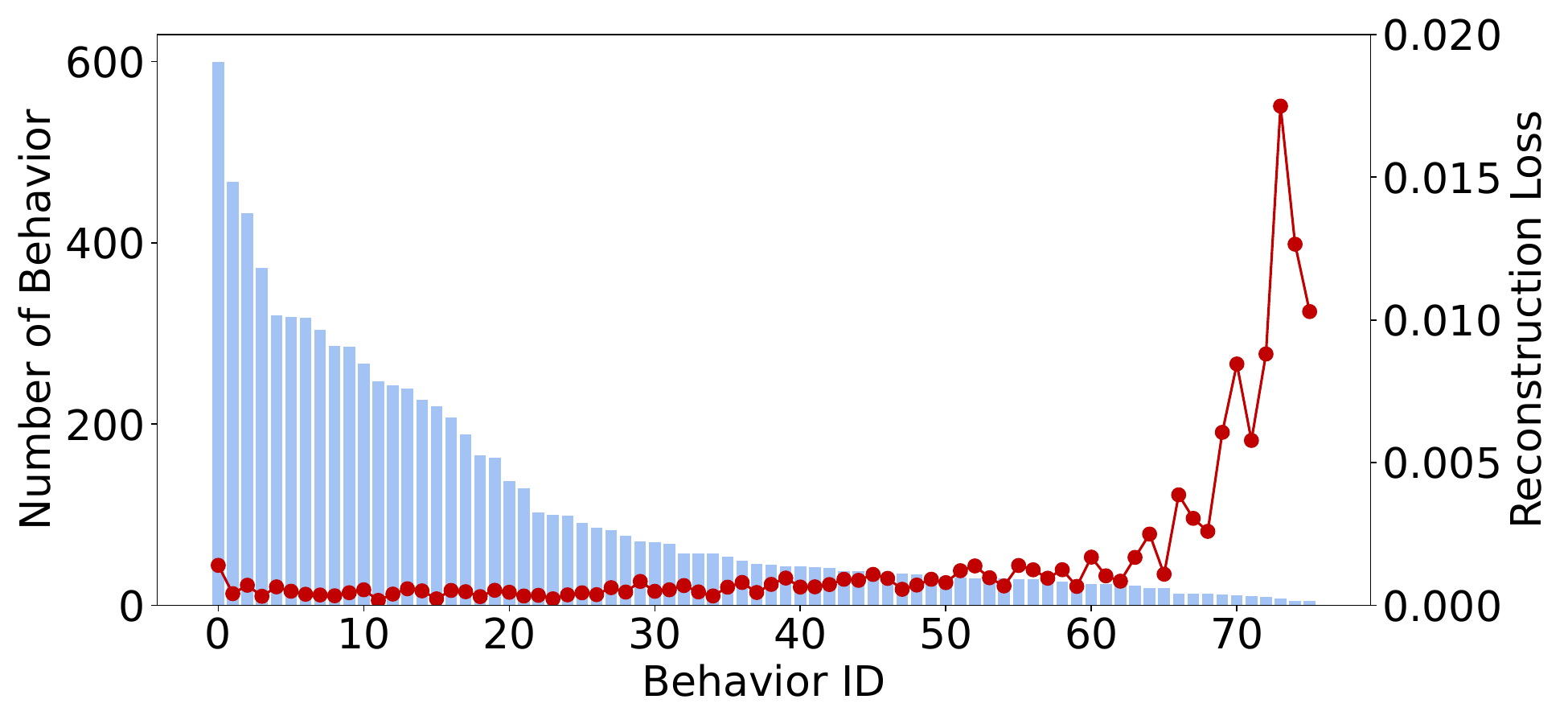}
\caption{Reconstruction losses for behaviors with different occurrence frequencies.}
\label{fig:unbalance}
\end{figure}

First, the occurrence frequencies of different user behaviors may be imbalanced, leading to challenges in learning the semantics of these behaviors. 
This user \textbf{behavior imbalance} can be attributed to individuals' living habits. For example, cook-related behaviors (e.g., using microwave and oven) of office workers may be infrequent, because they dine at their workplace on weekdays and only cook on weekends. On the other hand, some daily behaviors like turning on lights and watching TV of the same users can be more frequent.  Behavior imbalance complicates the learning process for models: some behaviors, which occur frequently in similar contexts, can be easily inferred, while others that rarely appear or manifest in diverse contexts can be more challenging to infer. We train an autoencoder model on AN dataset (shown in Table~\ref{tab:dataset}), record the occurrences and reconstruction loss of different behaviors. As shown in Figure~\ref{fig:unbalance}, with the number of occurrences of behavior decreases, the reconstruction loss tends to increase.

\begin{figure}[ht]
\centering
\includegraphics[width = .46\textwidth]{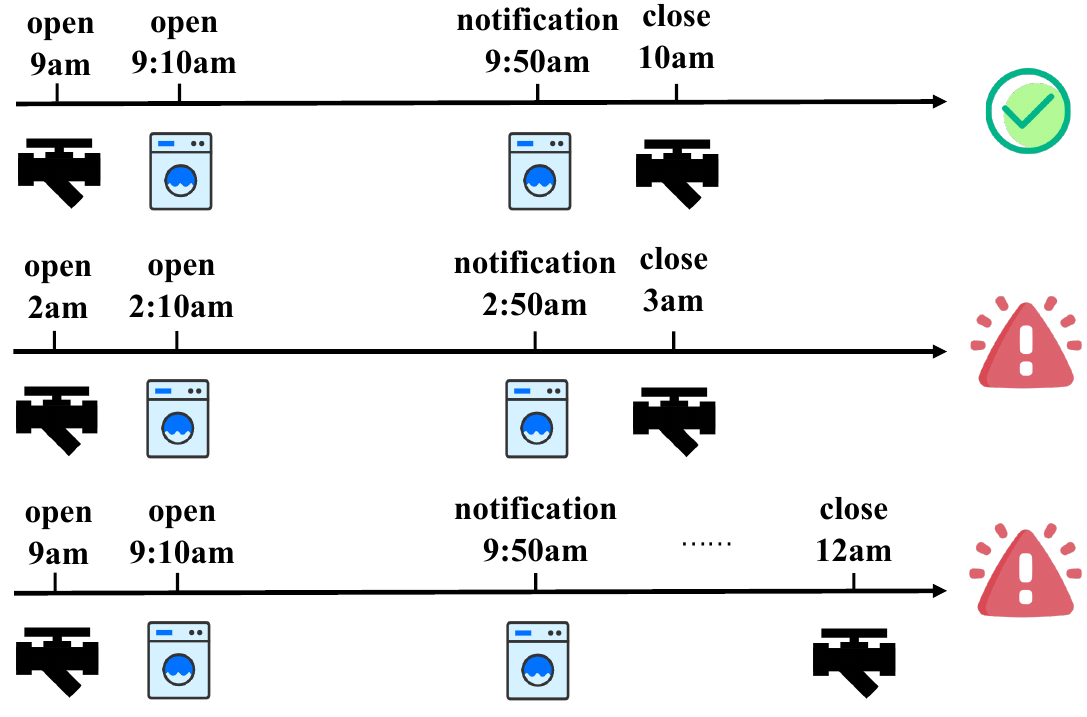}
\caption{Example of three user behaviors with the same behavior order. Sequence 2 and 3 are abnormal due to their inappropriate timing and excessive duration.}
\label{fig:time}
\end{figure}

Second, \textbf{temporal context}, e.g., the timing and duration of user behaviors, plays a significant role in abnormal behavior detection but is overlooked by existing solutions. For example, turning on the cooling mode of the air conditioner in winter is abnormal, but is normal in summer. Showering for 30-40 minutes is normal, but exceeding 2 hour suggests a user accident. Ignoring timing information hinders the identification of abnormal behavior patterns. As shown in Figure~\ref{fig:time}, sequence 1 represents a user's normal laundry-related behaviors. Sequences 2 and 3 follow the same order as sequence 1. However, in sequence 2, the water valve were opens at 2 o'clock in the night. In sequence 3, the duration between opening and closing the water valve is excessively long. Therefore these two sequences should be identified as abnormal behaviors possibly conducted by attackers intending to induce water leakage.

\begin{figure}[ht]
\centering
\includegraphics[width = .42\textwidth]{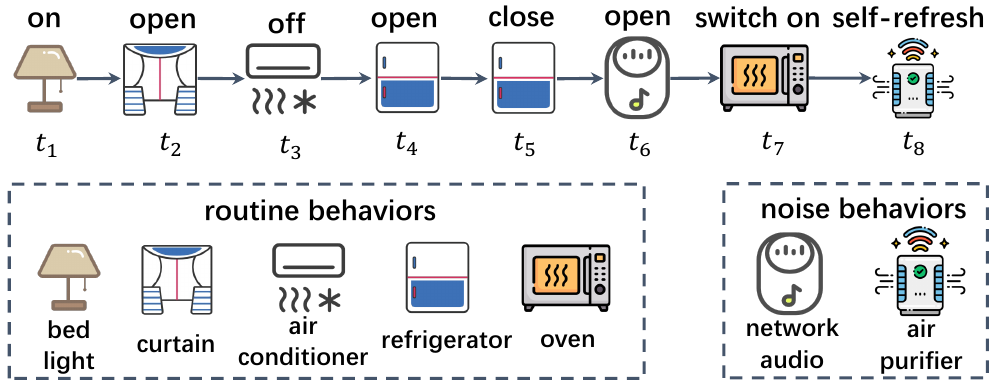}
\caption{An example of noise behaviors.}
\label{fig:noise}
\end{figure}

Third, arbitrary intents and passive device actions can cause \textbf{noise behaviors} in user behavior sequences, which interfere model's inference. 
Figure~\ref{fig:noise} shows noise behaviors in a behavior sequence related to a user's behaviors after getting up.  The user do some routine behaviors like ``turn on the bed light'', ``open the curtains'', ``switch off the air conditioner'', ``open the refrigerator'', ``close the refrigerator'' and ``switch on the oven''. However, there are also some sporadic actions which are not tightly related to the behavior sequence, including 1) active behaviors, e.g., suddenly deciding to ``turn on the network audio'' to listen to music; 2) passive behavior from devices, e.g., the ``self-refresh'' of the air purifier. These noise behaviors may also occur in other sequences with varying patterns. 
These noise behaviors introduces uncertainty that can disrupt the learning process and lead the model to misclassify sequences containing noise behaviors as anomalies. Therefore, treating noise behaviors on par with normal behaviors could potentially harm the model's performance, leading to increased losses.






In this paper, we propose \schemename \ to solve above challenges. \schemename \ is an autoencoder-based architecture, which learns to reconstruct normal behavior sequences during training and identify the behavior sequences with high reconstruction loss as anomaly. Firstly, we devise a \textbf{L}oss-guided \textbf{D}ynamic \textbf{M}ask \textbf{S}trategy (\textbf{LDMS}) to promote the model's learning of infrequent \textit{hard-to-learn} behaviors. Secondly, we introduce a \textbf{T}hree-level \textbf{T}ime-aware \textbf{P}osition \textbf{E}mbedding (\textbf{TTPE}) to integrate temporal information into positional embedding for detecting temporal context anomalies. Lastly, we propose a \textbf{N}oise-aware \textbf{W}eighted \textbf{R}econstruction \textbf{L}oss (\textbf{NWRL}) to assign distinct weights to routine behaviors and noise behaviors, thereby mitigating the impact of noise behaviors. Our codes are released to the GitHub \footnote{https://github.com/xjywhu/SmartGuard}. Our contributions can be summarized as follows:

\begin{itemize}

    \item We design \textbf{LDMS} to mask the behaviors with high reconstruction loss, thus encouraging the model to learn these \textit{hard-to-learn} behaviors.
    \item We propose \textbf{TTPE} to consider the order-level, moment-level and duration-level information of user behaviors meanwhile.
    \item We design \textbf{NWRL} to treat noisy behaviors and normal behaviors differently for learning robust behavior representations.  
\end{itemize}

\section{Related Work}

\subsection{User Behavior Modeling in Smart Homes}
Some works propose to model user behavior (i.e., user device interaction) based on deep learning.
~\cite{iotgaze} uses event transition graph to model IoT context and detect anomalies. In~\cite{dsnGraph}, authors build device interaction graph to learn the device state transition relationship caused by user actions. ~\cite{hawatcher} detects anomalies through correlational analysis of device actions and physical environment. ~\cite{DBLP:conf/huc/SrinivasanSW08} infers user behavior through readings from various sensors installed in the user's home.
IoTBeholder \cite{zou2023iotbeholder} utilizes attention-based LSTM to predict the user behavior from history sequences. SmartSense \cite{jeon2022accurate} leverages query-based transformer to model contextual information of user behavior sequences. DeepUDI \cite{xiao2023user} and SmartUDI \cite{xiao2023know} use relational gated graph neural networks, capsule neural networks and contrastive learning to model users' routines, intents and multi-level periodicities. However, above methods aim at predicting next behavior of user accurately, they can not be applied into abnormal behavior detection.

\subsection{Attacks and Defenses in Smart Homes}
An increasing number of attack vectors have been identified in smart homes in recent years. In addition to cyber attacks, it is also a concerning factor that IoT devices are often close association with the user's physical environment and they have the ability to alter physical environment. In this context, the automation introduces more serious security risks. Prior research has revealed that adversaries can leak personal information, and gain physical access to the home~\cite{ContexloT, DBLP:conf/uss/CelikBSATMU18}. In~\cite{spoof}, spoof attack is employed to exploit automation rules and trigger unexpected device actions. ~\cite{delay-sp, delay-dsn} apply delay-based attacks to disrupt cross-platform IoT information exchanges, resulting in unexpected interactions, rendering IoT devices and smart homes in an insecure state. This series of attacks aim at causing smart home devices to exhibit expected actions, thereby posing significant security threats. Therefore, designing an effective mechanism to detect such attacks is necessary. 6thSense \cite{sikder20176thsense} utilizes Naive Bayes to detect malicious behavior associated with sensors in smart homes. Aegis \cite{Siker19Aegis} utilizes a Markov Chain to detect malicious behaviors. ARGUS \cite{Rieger23ARGUS} designed an Autoencoder based on Gated Recurrent Units (GRU) to detect infiltration attacks. However, these methods ignore the behavior imbalance, temporal information and noise behaviors.

\section{Problem Formulation}
\label{sec:pf}

Let $\mathcal{D}$ denote a set of devices, $\mathcal{C}$ denote a set of device controls and $\mathcal{S}$ denote a set of behavior sequences.

\begin{myDef}   
(\textbf{Behavior}) A behavior $b=(t, d, c)$, is a 3-tuple consisting of time stamp $t$, device $d \in \mathcal{D}$ and device control $c \in \mathcal{C}$. 
\end{myDef}
For example, behavior \textit{b = (2022-08-04 18:30, air conditioner, air conditioner:switch on)} describes the behavior ``\textit{swich on the air conditioner}'' at 18:30 on 2022-08-04.


\begin{myDef}   
(\textbf{Behavior Sequence}) A behavior sequence $s=[b_{1}, b_{2}, \cdots, b_{n}] \in \mathcal{S}$ is a list of behaviors,ordered by their timestamps, and $n$ is the length of $s$.
\end{myDef}

We define the User Behavior Sequence (UBS) anomaly detection problem as follows.
\begin{myPro}
(\textbf{\problem \ Anomaly Detection}) Given a behavior sequence $s$, determine whether $s$ is an anomaly event or a normal event.
\end{myPro}



\begin{figure}[ht]
\centering
\includegraphics[width = .46\textwidth]{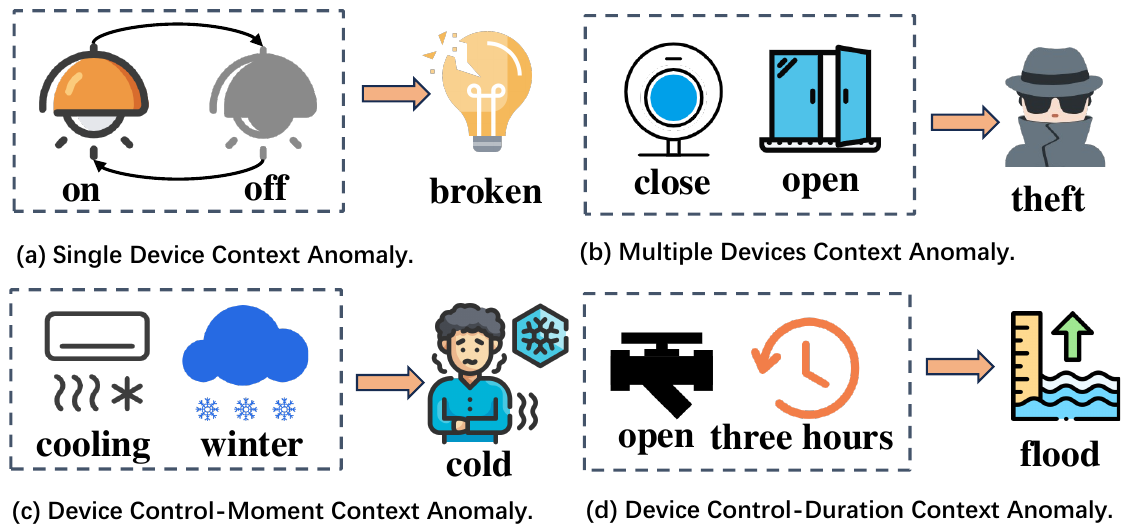}
\caption{Different types of anomaly behaviors.}
\label{fig:type}
\end{figure}

In this paper, we consider four types of abnormal behaviors:
\begin{itemize}
    \item \textbf{(SD)} Single Device context anomaly (Figure~\ref{fig:type}(a)), defined as unusual high frequency operations on a single device, e.g., frequently switching light on and off to break the light.
    \item \textbf{(MD)} Multiple Devices context anomaly (Figure~\ref{fig:type}(b)), defined as the simultaneous occurrence of behaviors on multiple devices that are not supposed to occur in the same sequence, e.g., turning off the camera and opening the window for burglary.
    \item \textbf{(DM)} Device control-Moment context anomaly (Figure~\ref{fig:type}(c)), defined as a device control occurring at an inappropriate time, e.g., turning on the cooling mode of an air conditioner in winter, potentially causing the user to catch a cold.
    \item \textbf{(DD)} Device control-Duration context anomaly (Figure~\ref{fig:type}(d)), defined as device controls that last for an inappropriate duration, e.g., leaving a water valve open for 3 hours for flood attack.
\end{itemize}

\begin{figure*}[ht]
\centering
\includegraphics[width = .95\textwidth]{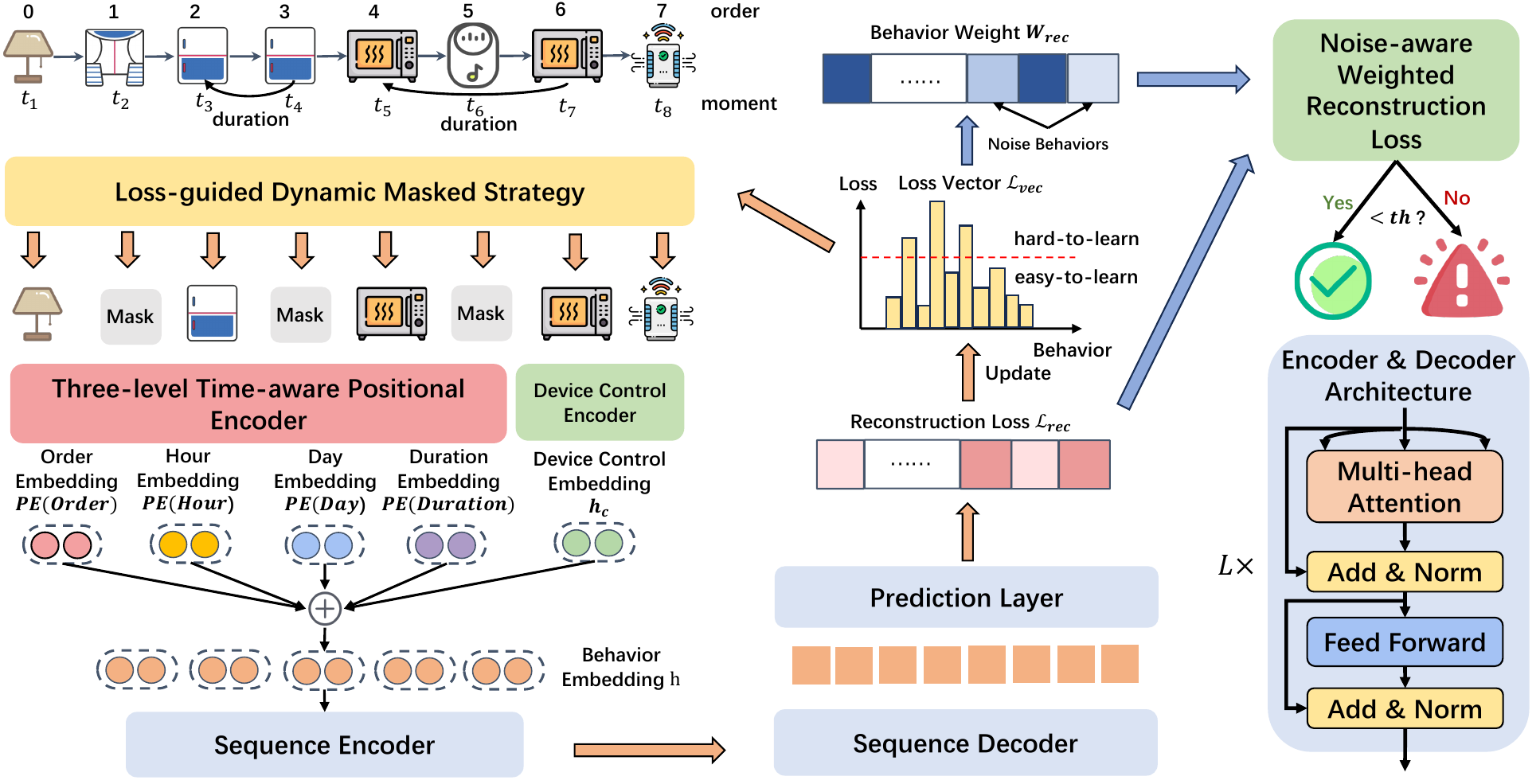}
\caption{The overview of \ \schemename.}
\label{fig:overview}
\end{figure*}

\section{Methodology}
\subsection{Solution Overview}



To achieve accurate user behavior sequence anomaly detection in smart homes, we propose \schemename, depicted in Figure~\ref{fig:overview}. The workflow of \schemename \ can be summarized as follows. During training, the Loss-guided Dynamic Mask Strategy (\S \ref{subsec:LDMS}) is initially employed to mask \textit{hard-to-learn} behaviors based on the loss vector $\mathcal{L}_{\text{vec}}$ from the previous epoch. Subsequently, the Three-level Time-aware Positional Encoder  (\S \ref{subsec:TTPE}) is applied to capture order-level, moment-level, and duration-level temporal information of the behaviors, producing the positional embedding $\overline{PE}$. This embedding is then added to the device control embedding $h_c$ to form the behavior embedding $\mathbf{h}$. Finally, $\mathbf{h}$ is fed into an $L$-layer attention-based encoder and decoder to extract contextual information for reconstructing the source sequence. During the inference phase, the Noise-aware Weighted Reconstruction Loss Noise-aware Weighted Reconstruction Loss (\S \ref{subsec:NWRL}) is utilized to assign different weights to various behaviors, determined by the loss vector from the training dataset, resulting in the final reconstruction loss $score$. If the $score$ surpasses the threshold $th$, \schemename \ triggers an alarm.



\subsection{Loss-guided Dynamic Mask Strategy}
\label{subsec:LDMS}
Autoencoders \cite{zhai2018autoencoder}, which take complete data instances as input and target to reconstruct the entire input data, are widely used in anomaly detection. Different from traditional autoencoders, masked autoencoders randomly mask a portion of input data, encoding the partially-masked data and aiming to reconstruct the masked tokens. By introducing a more meaningful self-supervised task, masked autoencoders have recently excelled in images \cite{he2022masked} learning. However, such reconstruction tasks without mask and with random mask are sub-optimal in our scenarios because they do not emphasize the learning of \textit{hard-to-learn} behaviors that occur rarely.

\begin{figure}[ht]
    \subfigure[Mean of reconstruction loss.]{
    \label{fig:loss}
    \centering
    \includegraphics[width = .2\textwidth]{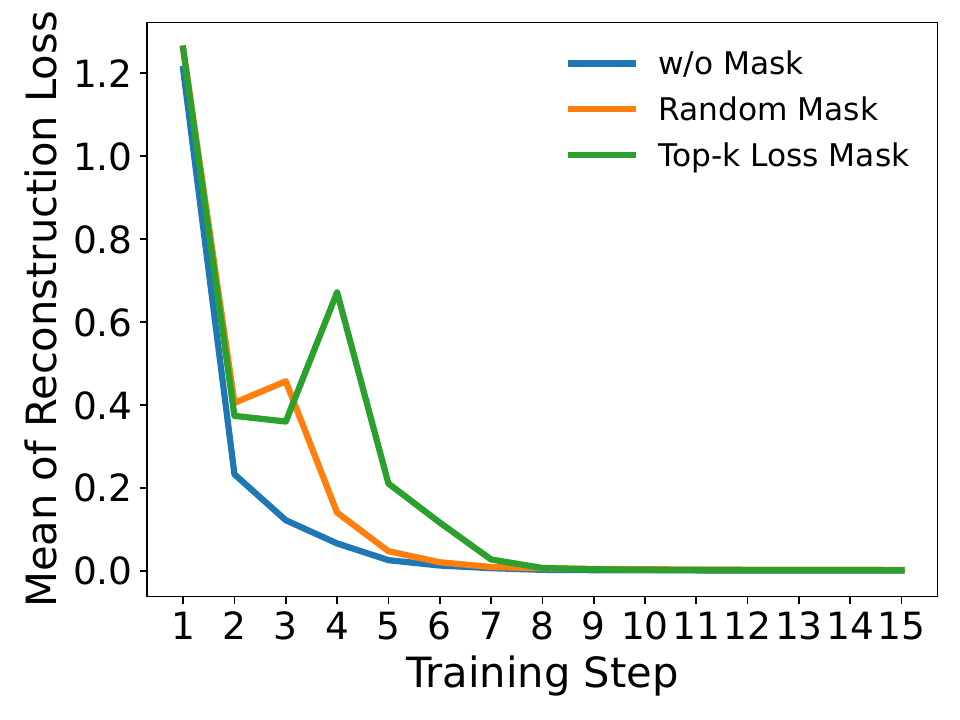}
    }
    \subfigure[Variance of reconstruction loss.]{
    \label{fig:variance}
    \centering
    \includegraphics[width = .2\textwidth]{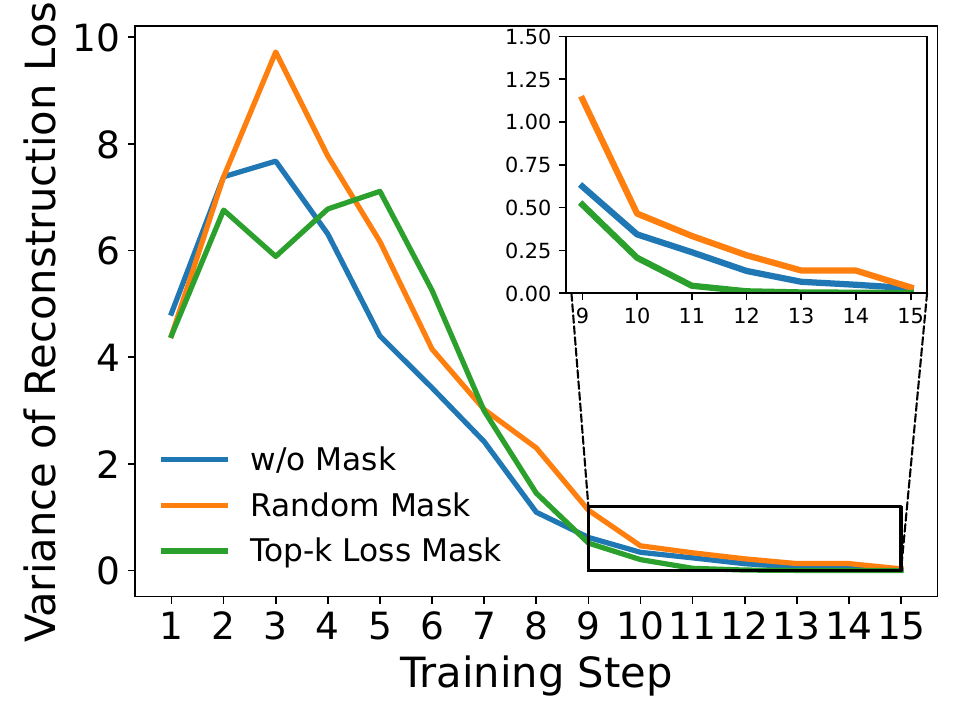}
    }
    \caption{Autoencoder training process on SP dataset under different mask strategies.}
    \label{fig:pre_ex}
\end{figure}

We conduct experiments to verify the performance of autoencoders trained with three mask options:
1) w/o mask: no mask strategy is used, the objective function is to reconstruct the input; 2) random mask: masking behaviors at every epoch randomly to reconstruct the masked behaviors; 3) top-$k$ loss mask: masking top $k$ behaviors with higher reconstruction loss to reconstruct the masked behaviors. We set mask ratio as 20\% for the latter two.
Figure~\ref{fig:pre_ex} shows the changing trends of the reconstruction loss and it's variance of different behavior during training on SP dataset (described in Table~\ref{tab:dataset}).
First, as shown in Figure~\ref{fig:loss}, the model without mask shows the fastest convergence trend, whereas the loss of the model with mask fluctuates. Model without mask can simultaneously learn all behaviors, facilitating rapid convergence. In contrast, the mask strategy only encourages the model to focus on learning masked behaviors, which may hinder initial-stage convergence. Second, the model with top-$k$ loss mask strategy shows lowest variance towards the end of training as shown in Figure~\ref{fig:variance}, because the top-$k$ loss mask strategy effectively encourages the model to learn \textit{hard-to-learn} behaviors (i.e., the behaviors with high reconstruction loss), thereby reducing the variance of behavior reconstruction losses.


In this paper, we design a Loss-guided Dynamic Mask Strategy. Intuitively, at the beginning of training, we encourage the model to learn the relatively easy task to \textbf{accelerate convergence}, i.e., behavior sequence reconstruction without mask. After training $N$ epochs without mask, we adopt the top-$k$ loss mask strategy to \textbf{encourage the model to learn the masked behaviors with high reconstruction loss}. We continuously track the model’s reconstruction loss of different behaviors by updating a loss vector in each epoch to guide the mask strategy in the next epoch. In epoch $ep$, the loss vector $\mathcal{L}^{ep}_{\text {vec}}$ is calculated as:
\begin{equation}
    \mathcal{L}^{ep}_{\text {vec}} = \left\{\ell_1, \ell_2, \ldots, \ell_c, \ldots, \ell_{|\mathcal{C}|}\right\}, c \in \mathcal{C},
\end{equation}
\begin{equation}
    \ell_c = \frac{1}{n_c} \sum_{i=1}^{n_{c}} \ell^{i}_c,
\end{equation}
where $n_c$ is the number of times the device control $c$ occurs in epoch $ep$, and $\ell_c$ is the average reconstruction loss of the device control $c$. 
In epoch $ep+1$, the mask vector for behavior sequence sample $s = [b_{1}, b_{2}, \cdots, b_{n}]$ is obtained as:
\begin{equation}
\label{equ:r}
mask(i)=\left\{\begin{array}{rl}
1, & \text{if} \ i \in sorted\_index[:\lfloor n \cdot r \rfloor] \\
0, & \text{if} \ i \notin sorted\_index[:\lfloor n \cdot r \rfloor]
\end{array}\right., i \in [1, n],
\end{equation}
\begin{equation}
    sorted\_index = \operatorname{argsort} \left(\left\{\mathcal{L}^{ep}_{\text {vec}}(b_1), \mathcal{L}^{ep}_{\text {vec}}(b_2), .... \mathcal{L}^{ep}_{\text {vec}}(b_n)\right\}\right),
\end{equation}
where $\operatorname{argsort}$ gets the sorted index with descending order of the elements in the vector, $r \in [0, 1]$ is the mask ratio, and $n$ is the length of behavior sequence $s$.

\subsection{Autoencoder with Temporal Information}
\subsubsection{\textbf{Three-level Time-aware Positional Encoder}}
\label{subsec:TTPE}


The temporal information in user behavior sequence data primarily resides in the timing of control behaviors, which can be examined from two perspectives: the absolute timing of each individual control behavior, and the relative timing gap between control actions on the same device. On the one hand, the relative timing gap between control actions on the same device reflects the duration the device is in some specific state and the operation frequency of the user. On the one hand, user behaviors are usually time-regulated, and the functionalities a device carried can determine the absolute timing users operate on it. For example, users usually operate lights in the morning and evening, and operate the microwave and the oven at meal time. Since certain operations frequently take place nearly simultaneously, we will also consider the order of behaviors to provide a more comprehensive characterization of behaviors that occur successively. Therefore, we incorporate three types of temporal information into our model. (1) \textbf{Order-level temporal information}: we use integer $order \in [0, n-1]$ to denotes the order-level information of the behavior, $n$ is the length of behaviors sequence $s$. (2) \textbf{Moment-level temporal information}: we represent the moment as hour of day $hour$ and day of week $day$ based on behavior's timestamp. (3) \textbf{Duration-level temporal information}: the duration for behavior $b$ is calculated as:




\begin{equation}
    duration_b = t(b)-t(b_{next})
\end{equation}
where $b$ and $b_{next}$ are the behaviors on the same device and $b_{next}$ is the first behavior after $b$ that operates on the device, and $t(b)$ represents the occurrence time of behavior $b$. 

Then, the positional embedding is calculated as:
\begin{equation}
    \begin{split}
    \overline{PE} &= w_{order} \cdot PE(pos) + w_{hour} \cdot PE(hour) + \\ 
    & w_{day} \cdot PE(day) + w_{dur} \cdot PE(duration),
\end{split}
\end{equation}
where $w_{order}$, $w_{hour}$, $w_{day}$ and $w_{dur}$ are learnable weights. $PE(\cdot)$ is a positional encoding function \cite{vaswani2017attention} defined as:
\begin{equation}
\label{equ:pos}
\begin{aligned}
PE_{(\cdot, 2i)} &=\sin \left(\cdot / 10000^{2 i / d} \right), \\
PE_{(\cdot, 2i+1)} &=\cos \left(\cdot / 10000^{2 i / d} \right),
\end{aligned}
\end{equation}
where $i$ denotes the $i$-th dimension of the positional embedding, $d$ is the dimension of temporal embedding.

To learn the representation $h_{c}$ for device control $c \in \mathcal{C}$, we first encode device control $c$ into a low-dimensional latent space through device control encoder, i.e., an embedding layer. Finally, we add positional embedding to the device control embedding as following to get the behavior embedding:
\begin{equation}
    \mathbf{h} = \overline{PE} + h_{c}.
\end{equation}


\subsubsection{\textbf{Sequence Encoder}}

To learn the sequence embedding, we employ transformer encoder \cite{vaswani2017attention} consisting of multi-head attention layer, residual connections and position-wise feed-forward network (FNN). Given an input behavior representation $\mathbf{h}$, the self-attention layer can effectively mine global semantic information of behavior sequence context by learning  query $\mathrm{Q}$, key $\mathrm{~K}$ and value $\mathrm{~V}$ matrices of different variables, which are calculated as:
\begin{equation}
\mathrm{Q}=\mathbf{h}\mathrm{W}^Q, \mathrm{~K}=\mathbf{h}\mathrm{W}^K, \mathrm{~V}=\mathbf{h}\mathrm{W}^V, 
\end{equation}
where $\mathrm{W}^Q, \mathrm{W}^K, \mathrm{W}^V$ are the transformation matrices. The attention score $\mathbf{A}$ is computed by:
\begin{equation}
\mathbf{A} = \operatorname{Attention}(Q, K, V)=\operatorname{softmax}\left(\frac{Q K^T}{\sqrt{d_k}}\right) V,
\end{equation}
where $d_{k}$ is the dimension of $K$. Multi-head attention is applied to improve the stability of the learning process and achieve higher performance. Then, the position-wise feed-forward network (FNN) and residual connections are adopted:


\begin{equation}
\mathbf{\overline{h}}=\operatorname{Trans}(\mathbf{h})=\mathbf{h}+\mathbf{Ah}+\mathrm{FNN}(\mathbf{h}+\mathbf{Ah}),
\end{equation}
where $\operatorname{Trans}(\cdot)$ is the transformer and $\operatorname{FNN}(\cdot)$ is a 2-layered position-wise feed-forward network \cite{vaswani2017attention}.

\subsubsection{\textbf{Sequence Decoder}} 
The decoder has the same architecture as the encoder. We input $\mathbf{\overline{h}}$ into the decoder to reconstruct the entire sequence, probabilities of target device controls are calculated as:
\begin{equation}
\mathbf{\widetilde{h_i}} = \operatorname{decoder}\left(\mathbf{\overline{h_i}}\right),
\end{equation}
\begin{equation}
\hat{\mathbf{y_i}}=\operatorname{softmax}\left(\mathbf{W}_{h}\mathbf{\widetilde{h_i}}\right),
\end{equation}where $\hat{\mathbf{y_i}}$ is the predicted probabilities of the $i$-th device control and $\mathbf{W}_{h} \in \mathbb{R}^{|\mathcal{C}| \times len(h)}$ is the learnable transformation matrix, $|\mathcal{C}|$ is the number of device controls, and $len(h)$ is the length of $h$.



\subsubsection{\textbf{Objective Function}}
We optimize the model to minimize the average reconstruction loss measured by cross-entropy loss:
\begin{equation}
\label{equ:obj}
\mathcal{L}_{rec}=\left\{\begin{array}{ll}
-\frac{1}{|\mathcal{S}|} \sum_{s \in \mathcal{S}} \sum^{|s|}_{i=1} \mathbf{y}_{i} \log \hat{\mathbf{y}}_{i}, & \text{if} \ epoch <= N \\
-\frac{1}{|\mathcal{S}|} \sum_{s \in \mathcal{S}} \sum^{|s|}_{i=1} mask_s(i) \mathbf{y}_{i} \log \hat{\mathbf{y}}_{i}, & \text{if} \ epoch > N 
\end{array}\right.,
\end{equation}
where $\mathcal{S}$ is the behavior sequences set, $|s|$ is the length of sequence $s$, $\mathbf{y}_i$ is the one-hot vector of the ground-truth label, $mask_s$ is the mask vector for sequence $s$, and $N$ is the training steps w/o mask.





\subsection{Noise-aware Weighted Reconstruction Loss}
\label{subsec:NWRL}

Although LDMS encourages the model to focus on learning behaviors with high reconstruction losses, it remains challenging to reconstruct noise behaviors due to their inherent uncertainty. The significant reconstruction loss associated with noise behaviors can overshadow other aspects during anomaly detection, potentially leading to the misclassification of normal sequences containing noise behaviors as anomalies.

To eliminate the interference of noise behaviors, we propose a Noise-aware Weighted Reconstruction Loss as the anomaly score. We can get the final loss vector after training:
\begin{equation}
    \mathcal{L}_{\text {vec}} = \left\{\ell_1, \ell_2, \ldots, \ell_c, \ldots, \ell_{|\mathcal{C}|}\right\}, c \in \mathcal{C},
\end{equation}
which is converted into the corresponding weight vector:
\begin{equation}
    \mathcal{W}_{vec}=\left\{w_1, w_2, \ldots, w_c, \ldots, w_{|C|}\right\}, w_k \in(0,1),
\end{equation}
by the following equation:
\begin{equation}
\label{equ:mu}
\mathcal{W}_{vec}=\operatorname{sigmoid}\left(-\frac{\operatorname{relu}(\mathcal{L}_{vec}-\mathbb{E}\left(\mathcal{L}_{vec}\right))}{\sqrt{\operatorname{Var}\left(\mathcal{L}_{vec}\right)} \cdot \mu}\right),
\end{equation}
where $\mu$ is a coefficient to adjust the input for sigmoid function, $\mathbb{E}$ and $\operatorname{Var}$ calculate the expectation and variance of the loss distribution, respectively. Relu function ensures that behaviors with losses less than $\mathbb{E}\left(\mathcal{L}_{vec}\right)$ (routine behaviors) are equally weighted. The sigmoid function assigns small weights to behaviors with high losses (potential noise behaviors). For each behavior $b_i$ in a sequence $s = \{b_1, b_2, \cdots, b_n\}$, we compute the weight $p_i$ as follows:
\begin{equation}
p_i=\frac{\mathcal{W}_{vec}(b_i)}{\sum_{j=1}^n \mathcal{W}_{vec}(b_j)}.
\end{equation}
Then, we can get the anomaly score of $s$ as the weighted sum of the reconstruction losses of behaviors in $s$:
\begin{equation}
score(s) = -\frac{1}{|s|}  \sum^{|s|}_{i=1} p_i \mathbf{y}_{i} \log \hat{\mathbf{y}}_{i}.
\end{equation}
\schemename \ can inference whether a behavior sequence $s_i$ is normal or abnormal based on the anomaly score:
\begin{equation}
s_i = \left\{\begin{array}{ll}
\text{Normal}, & \text{if} \ score(s_i) < th \\
\text{Abnormal}, & \text{if} \ score(s_i) > th
\end{array}\right.,
\end{equation}
where $th$ is the anomaly threshold. We take the 95\% quantile of the reconstruction loss distribution on the validation set as $th$.



\section{Experiments}
In this section, we conduct comprehensive experiments on three real-world datasets to answer the following key questions:

\begin{itemize}
    \item RQ1. \textbf{Performance.} Compared with other methods, does \schemename \ achieve better anomaly detection performance?
    \item RQ2. \textbf{Ablation study.} How will model performance change if we remove key modules of \schemename ?
    \item RQ3. \textbf{Parameter study.} How do key parameters affect the performance of \schemename ?
    \item RQ4. \textbf{Interpretability study.} Can \schemename \ give reasonable explanations for the detection results?
    \item RQ5. \textbf{Embedding space analysis.} Does \schemename \ successfully learn useful embeddings of behaviors and correct correlations between device controls and time?
\end{itemize}

\subsection{Experimental Setup}

\subsubsection{\textbf{Datasets}}

We train \schemename \ on three real-world datasets consisting of only normal samples, two (FR/SP) from public datasets\footnote{https://github.com/snudatalab/SmartSense} and one anonymous dataset (AN) collected by ourselves. The datasets description is shown in Table \ref{tab:dataset}. All datasets are split into training, validation and testing sets with a ratio of 7:1:2. To evaluate the performance of \schemename, we construct ten categories of abnormal behaviors as shown in Table~\ref{tab:types} and insert them among normal behaviors for simulating real anomaly scenarios.

\begin{table}[]
\caption{Datasets Description.}
\setlength{\tabcolsep}{.05em}{
\label{tab:dataset}
\resizebox{\linewidth}{!}{
\begin{tabular}{@{}ccccc@{}}
\toprule
\multicolumn{1}{c}{Name}  & \multicolumn{1}{c}{Time period (Y-M-D)} & \multicolumn{1}{c}{Sizes} & \multicolumn{1}{c}{\# Devices} & \multicolumn{1}{c}{\# Device controls} \\ \midrule
AN                                             & 2022-07-31$\sim$2022-08-31         & 1,765                         & 36                          & 141                                 \\
FR                                            & 2022-02-27$\sim$2022-03-25         & 4,423                         & 33                          & 222                                 \\
SP                                             & 2022-02-28$\sim$2022-03-30         & 15,665                        & 34                          & 234                                 \\
\bottomrule
\end{tabular}}}
\end{table}

\begin{table}[]
\caption{Anomaly Behaviors.}
\label{tab:types}
\begin{tabular}{@{}ll|ll@{}}
\toprule
Anomaly                                                                      & Type     & Anomaly                                                                       & Type     \\ \midrule
Light flickering                                                             & SD   & \begin{tabular}[c]{@{}l@{}}Open the airconditioner's \\ cool mode in winter\end{tabular} & DM   \\ \midrule
Camera flickering                                                            & SD   & \begin{tabular}[c]{@{}l@{}}Open the window\\ at midnight\end{tabular}           & DM   \\ \midrule
TV flickering                                                                & SD   & \begin{tabular}[c]{@{}l@{}} Open the watervalve \\ at midnight\end{tabular}        & DM   \\ \midrule
\begin{tabular}[c]{@{}l@{}}Open the window \\ while smartlock lock\end{tabular} & MD & Shower for long time                                                          & DD \\ \midrule
\begin{tabular}[c]{@{}l@{}}Close the camera \\ while smartlock lock\end{tabular}  & MD & \begin{tabular}[c]{@{}l@{}}Microwave runs \\ for long time\end{tabular}        & DD \\ \bottomrule
\end{tabular}
\end{table}

\subsubsection{\textbf{Baselines}}
We compare \schemename \ with existing general unsupervised anomaly detection methods and unsupervised anomaly behaviors detection methods in smart homes:

\begin{itemize}
    \item \textbf{Local Outiler Factor (LOF)} \cite{cheng2019outlier} calculates the density ratio between each sample and its neighbors to detect anomaly.
    \item \textbf{Isolation Forest (IF)} \cite{liu2008isolation} builds binary trees, and instances with short average path lengths are detected as anomaly.

    \item \textbf{6thSense} \cite{sikder20176thsense} utilizes Naive Bayes to detect malicious behavior associated with sensors in smart homes.
    \item \textbf{Aegis}\cite{Siker19Aegis} utilizes a Markov Chain-based machine learning technique to detect malicious behavior in smart homes.


    \item \textbf{OCSVM} \cite{amraoui2021ml} build a One-Class Support Vector Machine model to prevent malicious control of smart home systems.
    \item \textbf{Autoencoder} \cite{chen2018autoencoder} learns to reconstruct normal data and then uses the reconstruction error to determine whether the input data is abnormal.
    \item \textbf{ARGUS}\cite{Rieger23ARGUS} designed an Autoencoder based on Gated Recurrent Units (GRU) to detect IoT infiltration attacks.

    \item \textbf{TransformerAutoencoder (TransAE)} \cite{vaswani2017attention} uses self-attention mechanism in the encoder and decoder to achieve context-aware anomaly detection.
    
\end{itemize}

\subsubsection{\textbf{Evaluation metrics}}

We use common metrics such as \textit{False Positive rate}, \textit{False Negative Rate}, \textit{Recall}, and \textit{F1-Score} to evaluate the performance of \schemename.


\subsubsection{\textbf{Complexity analysis}}
Suppose the embedding size is $em$, and the behavior sequence length is $n$. The computational complexity of \schemename \ is mainly due to the self-attention layer and the feed-forward network, which is $O(n^2d + nd^2)$. The dominant term is typically $O(n^2d)$ from the self-attention layer. \schemename \ only takes 0.0145s, which shows that it can detect abnormal behaviors in real time.

\begin{table*}[]
\caption{Performance comparison on three real world datasets.}
\label{tab:com}
\setlength{\tabcolsep}{0.5em}
\begin{tabular}{@{}cccccccccccc@{}}
\toprule
Dataset             & Type                & Metric   & \multicolumn{1}{c}{LOF} & \multicolumn{1}{c}{IF} & \multicolumn{1}{c}{6thSense} & \multicolumn{1}{c}{Aegis} & \multicolumn{1}{c}{OCSVM} & \multicolumn{1}{c}{Autoencoder} & \multicolumn{1}{c}{ARGUS} & \multicolumn{1}{c}{TransAE} & \multicolumn{1}{c}{SmartGuard} \\ \midrule
\multirow{8}{*}{AN} & \multirow{2}{*}{SD} & Recall   & 0.0275                  & 0.4105                              & 0.4680                   & 0.2902                      & 0.5399                    & 0.9832                          & 0.9858     & \underline{0.9882}              & \textbf{0.9986}                                \\
                    &                     & F1 Score & 0.0519                  & 0.4972                              & 0.5196                  & 0.3672                    & 0.5862                    & 0.9915                          & \underline{0.9928}     & 0.9908               & \textbf{0.9967}                               \\ \cmidrule(l){2-12}
                    & \multirow{2}{*}{MD} & Recall   & 0.0745                  & 0.4039                              & 0.5941                  & 0.4431                      & 0.6039                    & 0.5156                          & 0.5666         & \underline{0.6216}           & \textbf{0.9745}                               \\
                    &                     & F1 Score & 0.1357                  & 0.4824                              & 0.6215                  & 0.4718                    & 0.6553                    & 0.6692                          & 0.7135    & \underline{0.7557}                & \textbf{0.9832}                               \\ \cmidrule(l){2-12}
                    & \multirow{2}{*}{DM} & Recall   & 0.0784                  & 0.4373                              & 0.3745                  & 0.5647                      & 0.3510                     & 0.5196                          & 0.5313         & \underline{0.6078}           & \textbf{0.9961}                               \\
                    &                     & F1 Score & 0.1418                  & 0.5174                              & 0.4817                  & 0.5647                    & 0.4257                    & 0.6725                          & 0.6843         & \underline{0.7452}           & \textbf{0.9941}                               \\ \cmidrule(l){2-12}
                    & \multirow{2}{*}{DD} & Recall   & 0.0961                  & 0.3451                              & 0.1980                   & \underline{0.7804}                      & 0.4961                    & 0.5137                          & 0.5117       & 0.5294             & \textbf{0.9980}                               \\
                    &                     & F1 Score & 0.1713                  & 0.4282                              & 0.3108                  & \underline{0.7044}                    & 0.5967                    & 0.6675                          & 0.6675         & 0.6818           & \textbf{0.9951}                               \\ \cmidrule(l){1-12} 
\multirow{8}{*}{FR} & \multirow{2}{*}{SD} & Recall   & 0.3541                  & 0.2444                              & 0.2907                  & 0.3915                 & 0.5918                    & 0.9816                       & 0.9796         & \underline{0.9864}           & \textbf{0.9979}                               \\
                    &                     & F1 Score & 0.4804                  & 0.3655                              & 0.4167                  & 0.4542                 & 0.6612                    & 0.9907                          & 0.9897       & \underline{0.9921}             & \textbf{0.9932}                               \\ \cmidrule(l){2-12}
                    & \multirow{2}{*}{MD} & Recall   & 0.4275                  & 0.2980                               & 0.6567                  & 0.7098                 & 0.4384                    & 0.9726                          & \underline{0.9875}       & 0.9782            & \textbf{0.9984}                                \\
                    &                     & F1 Score & 0.5192                  & 0.4230                              & 0.6092                  & 0.3827                 & 0.5534                    & 0.9861                          & 0.9783       & \underline{0.9874}            & \textbf{0.9907}                               \\ \cmidrule(l){2-12}
                    & \multirow{2}{*}{DM} & Recall   & 0.3825                  & 0.3191                              & 0.5461                  & \underline{0.7619}                 & 0.3920                     & 0.4952                          & 0.6676       & 0.6529             & \textbf{0.9985}                               \\
                    &                     & F1 Score & 0.4830                   & 0.4494                              & 0.6124                  & 0.6822                  & 0.4940                     & 0.6508                          & \underline{0.7867}      & 0.7779             & \textbf{0.9912}                               \\ \cmidrule(l){2-12}
                    & \multirow{2}{*}{DD} & Recall   & 0.3572                  & 0.1850                               & 0.5358                  & \underline{0.9743}                 & 0.6267                    & 0.4397                          & 0.7329         & 0.6098           & \textbf{0.9981}                               \\
                    &                     & F1 Score & 0.4375                  & 0.2806                              & 0.5880                   & 0.4481                 & 0.6422                    & 0.6013                          & \underline{0.8382}        & 0.7479            & \textbf{0.9921}                               \\ \cmidrule(l){1-12} 
\multirow{8}{*}{SP} & \multirow{2}{*}{SD} & Recall   & 0.2197                  & 0.2643                              & 0.6979                  & 0.1618                      & 0.5332                    & \underline{0.9824}                          & 0.9795        & 0.9172            & \textbf{0.9862}                               \\
                    &                     & F1 Score & 0.3350                   & 0.3857                              & 0.7248                  & 0.2164                    & 0.6155                    & \textbf{0.9911}                          & \underline{0.9896}         & 0.9489           & 0.9831                               \\ \cmidrule(l){2-12}
                    & \multirow{2}{*}{MD} & Recall   & 0.2786                  & 0.3399                              & 0.6317                  & 0.7445                      & 0.3840                     & 0.5645                          & 0.9696       & \underline{0.9936}             & \textbf{0.9961}                               \\
                    &                     & F1 Score & 0.3916                  & 0.4632                              & 0.6440                   & 0.6636                    & 0.5026                    & 0.7095                          & \underline{0.9845}         & \textbf{0.9866}           & 0.9830                                \\ \cmidrule(l){2-12}
                    & \multirow{2}{*}{DM} & Recall   & 0.2780                   & 0.3465                              & 0.6080                   & \underline{0.8121}                      & 0.5351                    & 0.3074                          & 0.5297       & 0.5451             & \textbf{0.9198}                               \\
                    &                     & F1 Score & 0.4112                  & 0.4918                              & 0.6935                  & \underline{0.7758}                 & 0.6341                    & 0.4649                          & 0.6847             & 0.6962       & \textbf{0.9498}                               \\ \cmidrule(l){2-12}
                    & \multirow{2}{*}{DD} & Recall   & 0.2109                  & 0.1763                              & 0.5449                  & 0.8001                      & \underline{0.8293}                    & 0.6455                          & 0.6455        & 0.6456            & \textbf{0.9961}                               \\
                    &                     & F1 Score & 0.3052                  & 0.2627                              & 0.6343                  & 0.6545                 & 0.7311                    & \underline{0.7685}                          & 0.7658             & 0.7653       & \textbf{0.9788}                               \\ \cmidrule(l){1-12} 
\end{tabular}
\end{table*}


\subsection{Performance Comparison (RQ1)}
We use grid search to adjust the parameters of \schemename \ and report the overall performance of \schemename \ and all baselines in Table~\ref{tab:com}. Bold values indicate the optimal performance among all schemes, and underlined values indicate the second best performance. First, \schemename \ outperforms all competitors in most cases. This is because \schemename simultaneously considers the temporal information, behavior imbalance and noise behaviors. Second, \schemename\ significantly improves the performance on DM and DD type anomalies detection. We ascribe this superiority to our TTPE's effective mining of temporal information of behaviors. Third, the LOF, IF and 6thSense show the worst performance. Aegis and OCSVM outperforms LOF, IF and 6thSense, which benifits from the Markov Chain's modeling of behavior transitions and SVM’s powerful kernel function. The Autoencoder outperform the traditional models because of stronger sequence modeling capability. ARGUS outperforms Aueocoder because of stronger sequence modeling capability of GRU.  By exploiting transformer to mine contextual information, TransAE achieves better performance than all other baselines, but is still inferior to our proposed scheme.

\subsection{Ablation Study (RQ2)}

\begin{table}[]
\caption{The F1-Score of 5 variants ($C_0$-$C_4$) on AN dataset.}
\label{tab:ablation}
\setlength{\tabcolsep}{0.4em}
\begin{tabular}{@{}cccccccc@{}}
\toprule
LDMS                     & TTPE                     & NWRL                     & & SD     & MD     & DM     & DD     \\ \midrule
{\color[HTML]{FF0000} X} & {\color[HTML]{FF0000} X} & {\color[HTML]{FF0000} X} & $C_0$    & 0.9908 & 0.7557 & 0.7452  & 0.6818 \\
{\color[HTML]{00CD66} Y}                        & {\color[HTML]{00CD66} Y}                        & {\color[HTML]{FF0000} X} & $C_1$    & 0.9877 & 0.9708 & 0.9767 & 0.9817 \\
{\color[HTML]{00CD66} Y}                        & {\color[HTML]{FF0000} X} & {\color[HTML]{00CD66} Y}                        & $C_2$    & 0.9883 & 0.6716 & 0.6783 & 0.6799 \\
{\color[HTML]{FF0000} X} & {\color[HTML]{00CD66} Y}                        & {\color[HTML]{00CD66} Y}                       & $C_3$    & 0.9902 & 0.9766 & 0.9835 & 0.9855 \\
{\color[HTML]{00CD66} Y}                        & {\color[HTML]{00CD66} Y}                        & {\color[HTML]{00CD66} Y}                        & $C_4$    & \textbf{0.9967}  & \textbf{0.9832} & \textbf{0.9941} & \textbf{0.9951} \\ \bottomrule
\end{tabular}
\end{table}

\schemename \ mainly consists of three main components: \textbf{L}oss-guided \textbf{D}ynamic \textbf{M}ask \textbf{S}trategy (\textbf{LDMS}), \textbf{T}hree-level \textbf{T}ime-aware \textbf{P}osition \textbf{E}mbedding (\textbf{TTPE}) and \textbf{N}oise-aware \textbf{W}eighted \textbf{R}econstruction \textbf{L}oss (\textbf{NWRL}). To investigate different components’ effectiveness in \schemename, we implement 5 variants of \schemename 
\ for ablation study ($C_0$-$C_4$). {\color[HTML]{00CD66} Y} represents adding the corresponding components, {\color[HTML]{FF0000} X} represents removing the corresponding components. $C_4$ is \schemename \ with all three components.  As shown in Table~\ref{tab:ablation}, each component of \schemename \ has a positive impact on results. The combination of all components brings the
best results, which is much better than using any subset of the three components.

\subsection{Parameter Study (RQ3)}

\subsubsection{\textbf{The mask ratio $r$ and the training step $N$ without mask}} 
\label{subsec:mask}
\begin{figure}[ht]
    \subfigure[SD Anomaly.]{
    \label{fig:SD}
    \centering
    \includegraphics[width = .22\textwidth]{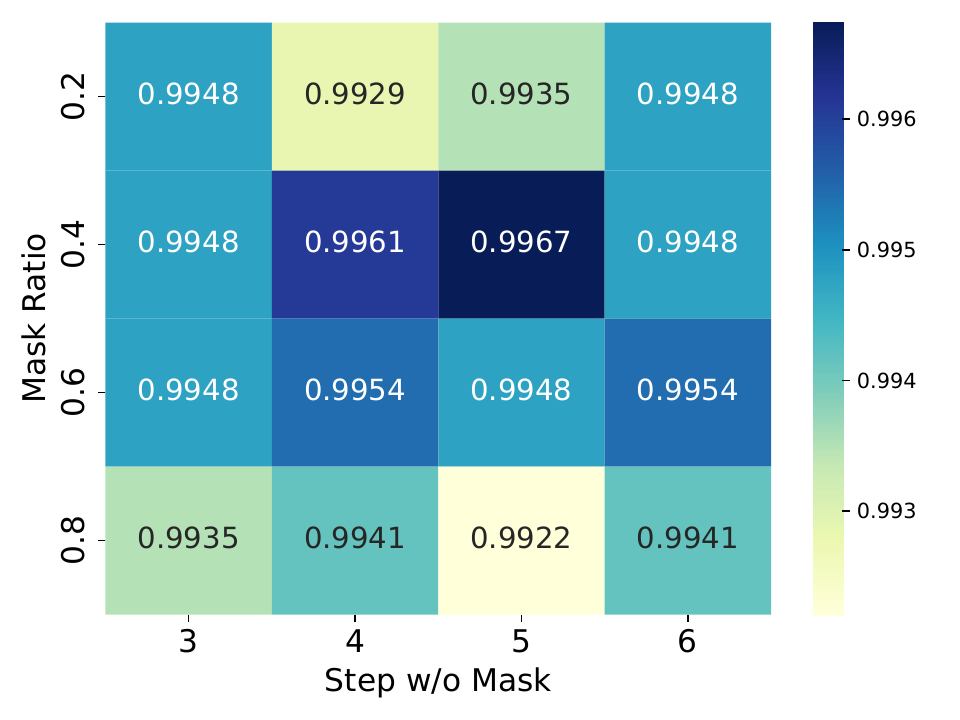}
    }
    \subfigure[MD Anomaly.]{
    \label{fig:MD}
    \centering
    \includegraphics[width = .22\textwidth]{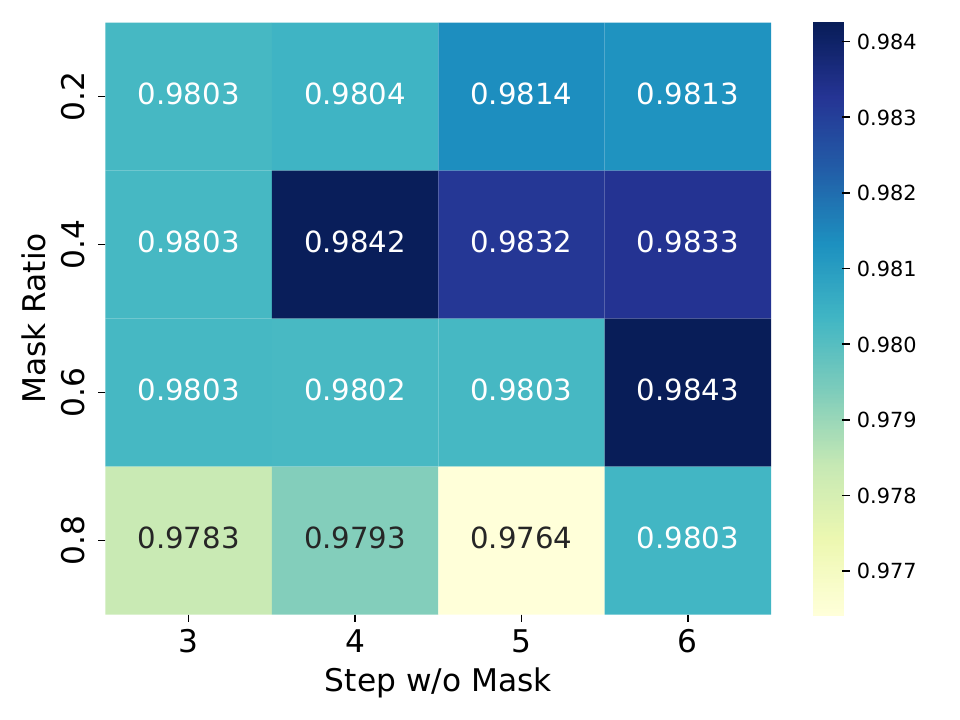}
    }
    \subfigure[DM Anomaly.]{
    \label{fig:DM}
    \centering
    \includegraphics[width = .22\textwidth]{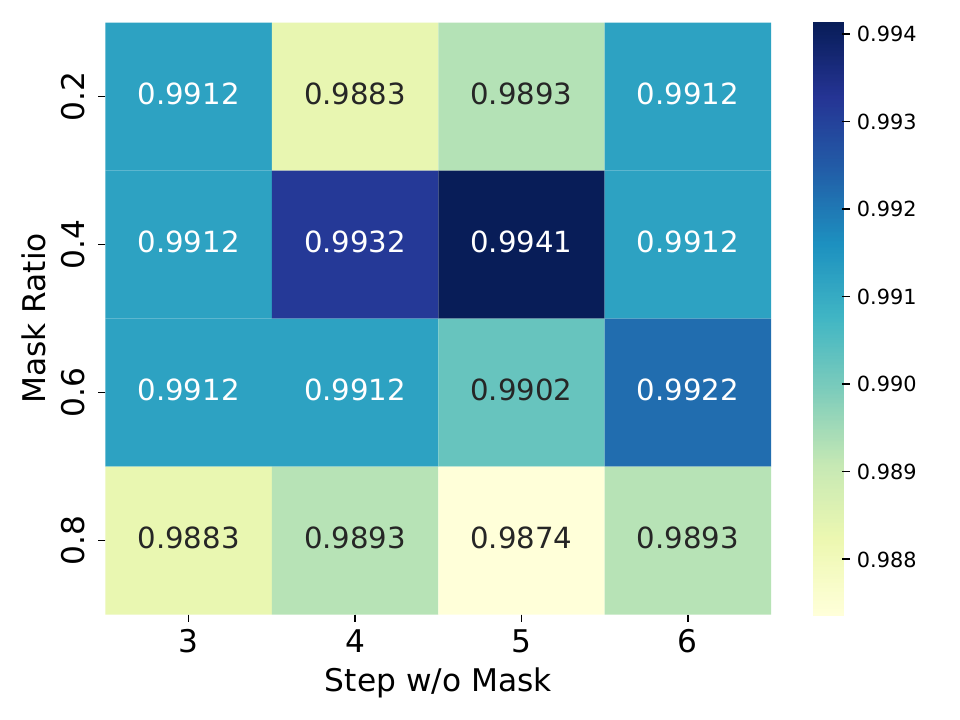}
    }
    \subfigure[DD Anomaly.]{
    \label{fig:DD}
    \centering
    \includegraphics[width = .22\textwidth]{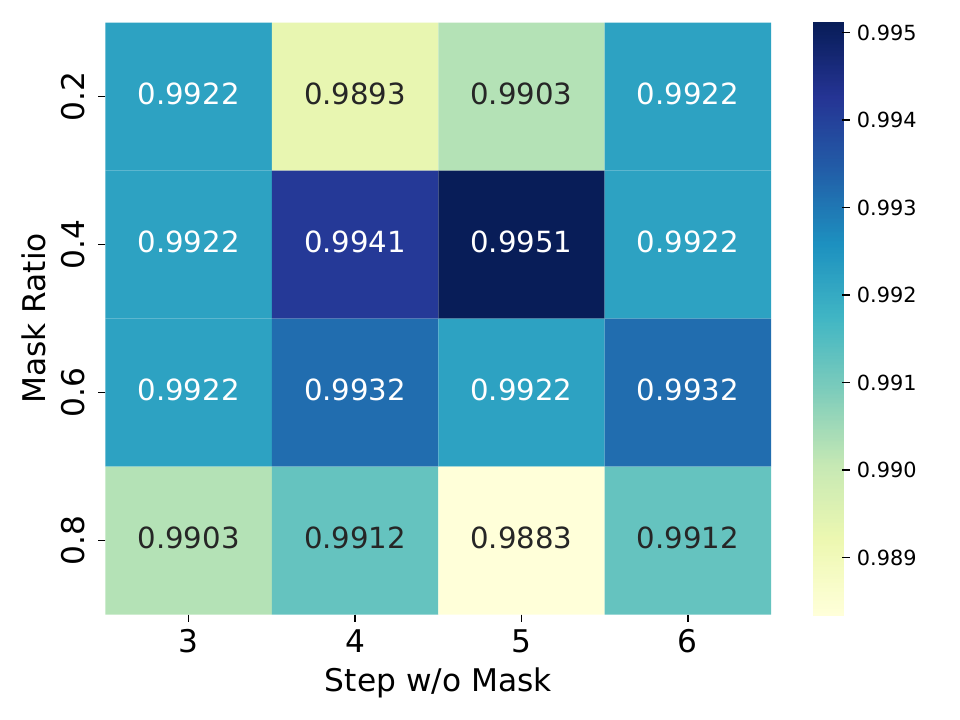}
    }
    \caption{Performance under different mask ratio and step w/o mask on AN dataset.}
    \label{fig:mask}
\end{figure}

Figure~\ref{fig:mask} illustrates that \schemename \ achieves the optimal performance when $r=0.4$ and $N=5$. The parameter $r$ (Equation~\ref{equ:r}) determines the difficulty of the model learning task. A smaller $r$ fails to effectively encourage the model to learn \textit{hard-to-learn} behaviors, while a larger $r$ increases the learning burden on the model, consequently diminishing performance. As for training steps without a mask, a smaller $N$ hinders the model from converging effectively at the beginning stage, whereas a larger $N$ impedes the model's ability to learn \textit{hard-to-learn} behaviors towards the end, resulting in degraded performance.


\subsubsection{\textbf{$\mu$ of Noise-aware Weighted Reconstruction Loss}} 
\begin{figure}[ht]
    \subfigure[False Positive Rate.]{
    \label{fig:fpr}
    \centering
    \includegraphics[width = .22\textwidth]{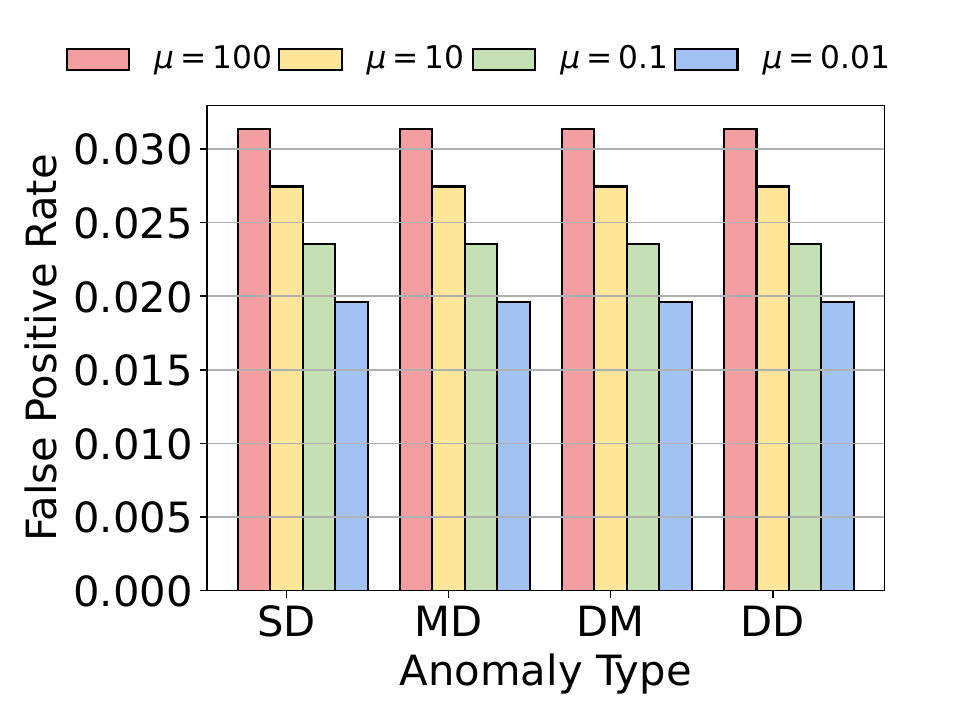}
    }
    \subfigure[False Negative Rate.]{
    \label{fig:fnr}
    \centering
    \includegraphics[width = .22\textwidth]{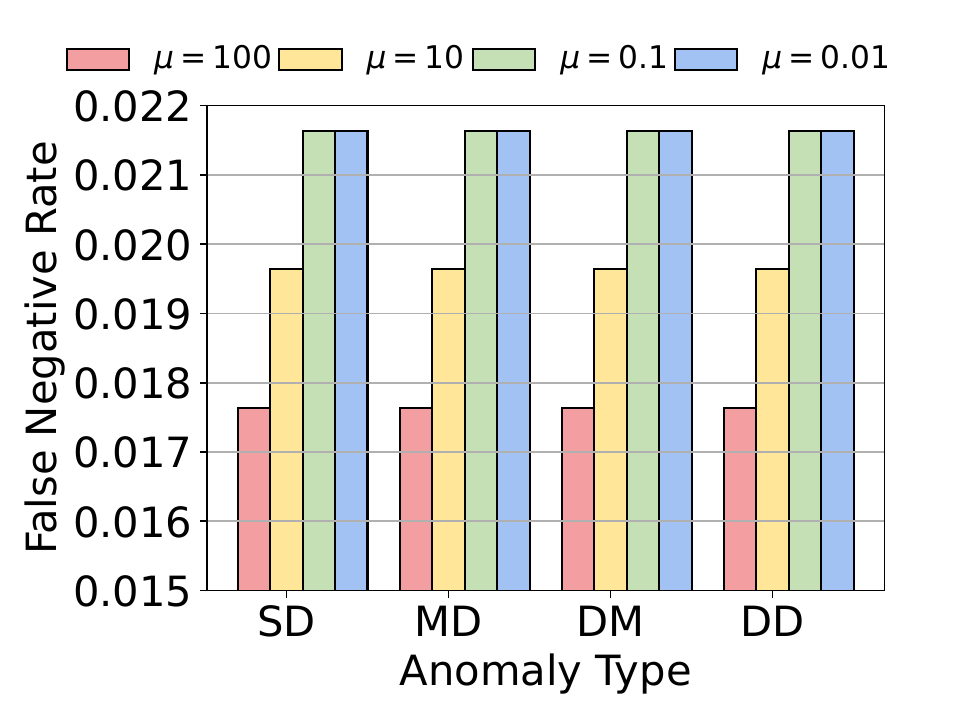}
    }
    \caption{False Positive Rate and False Negative Rate on AN dataset under different $\mu$.}
\end{figure}

The parameter $\mu$ 
 (Equation~\ref{equ:mu}) controls the weights assigned to potential noise behaviors. A smaller $\mu$ results in a smaller weight for noise behaviors, while a larger $\mu$ leads to a greater weight for noise behaviors. As illustrated in Figure~\ref{fig:fpr}, the False Positive Rate gradually decreases as $\mu$ decreases, benefiting from the reduced loss weight assigned to noise behaviors. However, as depicted in Figure~\ref{fig:fnr}, the False Negative Rate slightly increases as $\mu$ decreases. When $\mu=0.1$, \schemename \ achieves a balance, minimizing both the False Positive Rate and the False Negative Rate.


\subsubsection{\textbf{The embedding size $em$}} 
\begin{figure}[ht]
    \subfigure[Embedding dimension.]{
    \label{fig:em}
    \centering
    \includegraphics[width = .22\textwidth]{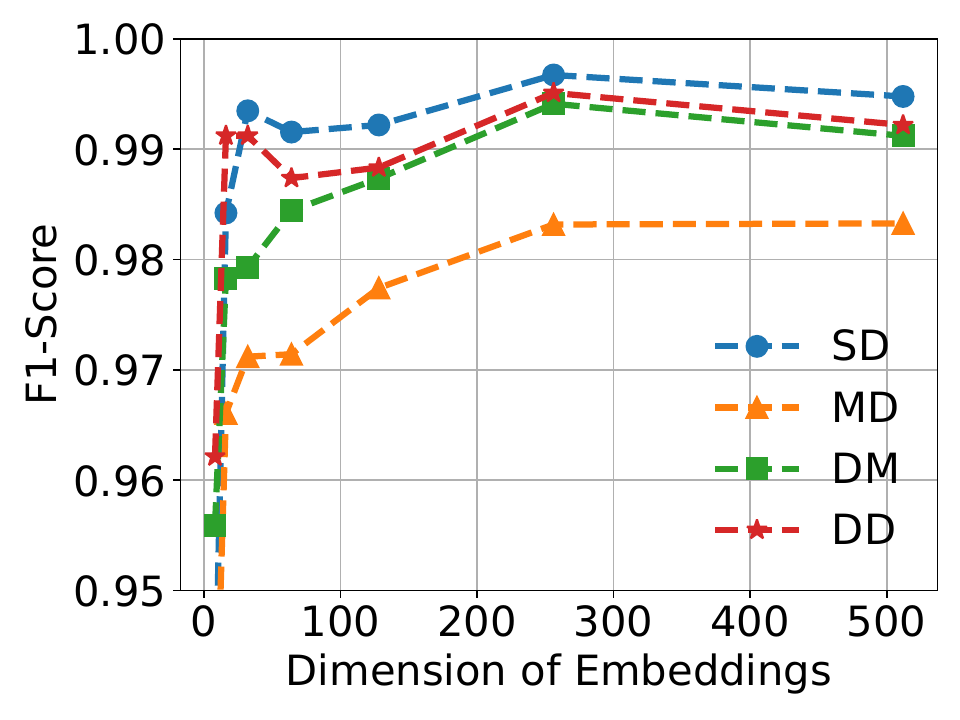}
    }
    \subfigure[Layers of encoder/decoder.]{
    \label{fig:layer}
    \centering
    \includegraphics[width = .22\textwidth]{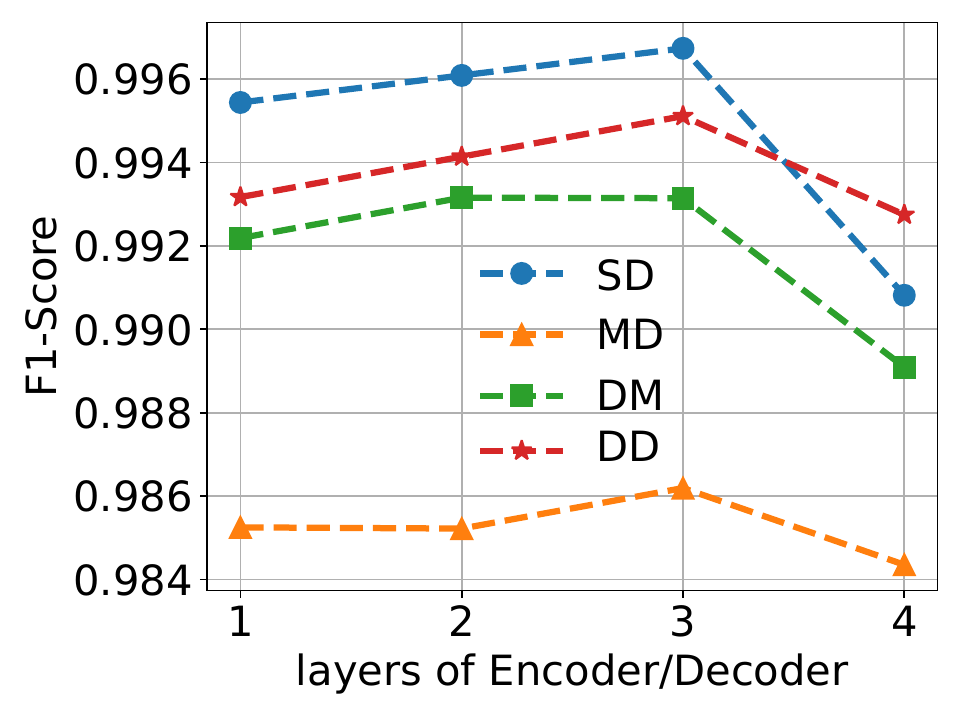}
    }
    \caption{The influence of embedding dimension and encoder/decoder layer number on AN dataset.}
\end{figure}

We fine-tune the embedding size for time and device control, ranging from 8 to 512. As depicted in Figure~\ref{fig:em}, an initial increase in the embedding dimension results in a notable performance improvement, which is attributed to the larger dimensionality enabling behavior embedding to capture more comprehensive information about the context, thereby furnishing valuable representations for other modules in \schemename. Nevertheless, excessively large sizes (e.g., > 256) can lead to performance degradation due to over-fitting.

\subsubsection{\textbf{Number of layers $L$ of encoder and decoder}} Figure~\ref{fig:layer} shows the performance of \schemename \ with different numbers of layers. When $L$ increases, F1-Score first increases and then decreases, reaching the optimal value at 3 layers, because fewer layers leads to under-fitting, and too many layers leads to over-fitting.

\subsection{Case Study (RQ4)}
\begin{figure}[ht]
\centering
\includegraphics[width = .45\textwidth]{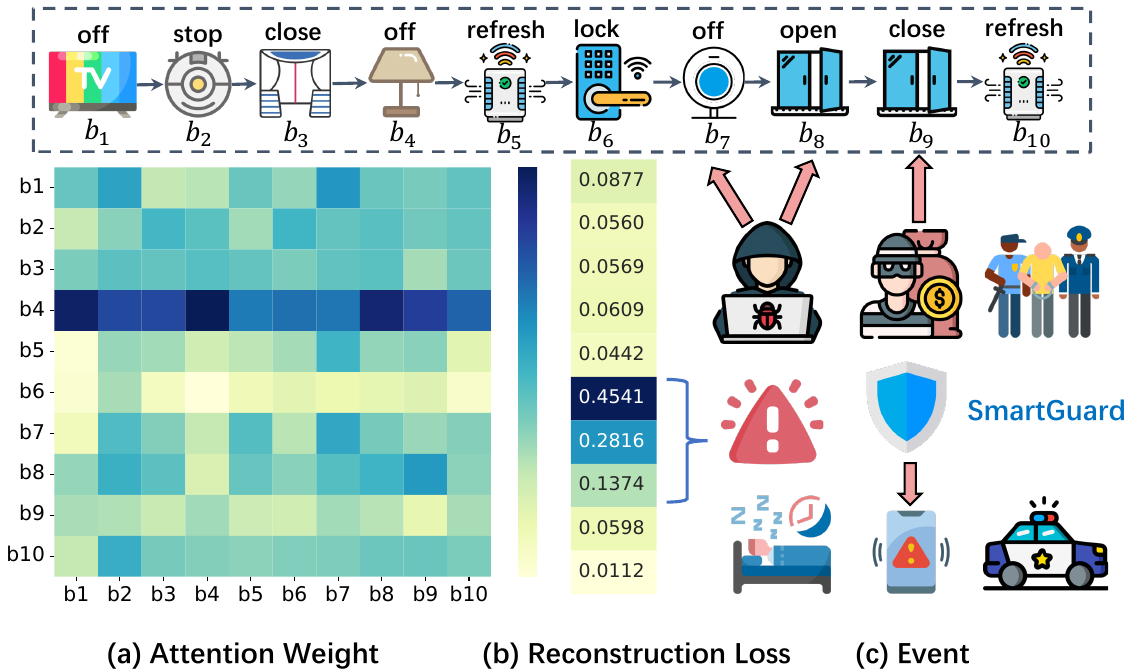}
\caption{(a) Attention weights, (b) reconstrution loss and (c) the corresponding events.}
\label{fig:case}
\end{figure}

To assess the interpretability of \schemename, we select a behavior sequence from the test set of the AN dataset and visualize its attention weights and reconstruction loss. Illustrated in Figure~\ref{fig:case}, the user initiated a sequence of actions: turning off the TV, stopping the sweeper, closing the curtains, switching off the bedlight, and locking the smart lock before going to sleep. Subsequently, an attacker took control of IoT devices, turning off the camera, and opening the window for potential theft. Examining Figure~\ref{fig:case}(a), we observe that the attention weights between behaviors $b_6$, $b_7$, $b_8$, and other behaviors in the sequence are relatively smaller. This suggests that $b_6$, $b_7$, $b_8$, and other behaviors lack contextual relevance and are likely abnormal. Turning to Figure~\ref{fig:case}(b), the reconstruction losses for behaviors $b_6$, $b_7$, and $b_8$ are notably high. \schemename \ identifies these anomalies in the sequence, triggering an immediate alarm.

\subsection{Embedding Space Analysis (RQ5)}

\begin{figure}[ht]
\centering
\includegraphics[width = .45\textwidth]{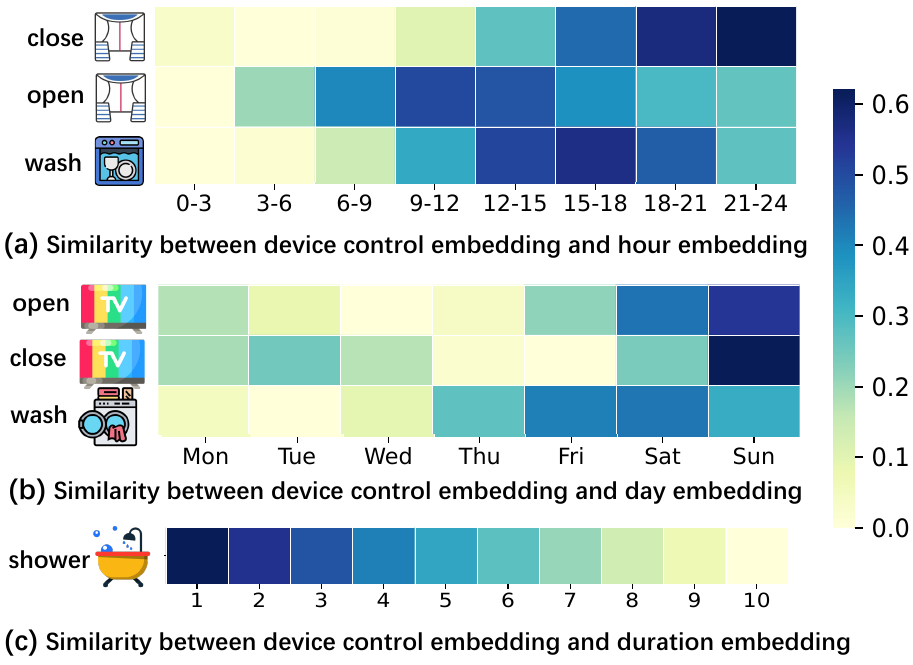}
\caption{Similarity between device control embedding and time embedding.}
\label{fig:es}
\end{figure}

We visualize the similarity between device embeddings and time embeddings (i.e., hour embedding, day embedding and duration embeddings) to analyze whether the model effectively learns the relationship between behaviors. As shown in Figure~\ref{fig:es}(a), opening the curtains usually occurs between 6-9 and 9-12 o'clock because users usually get up during this period, while closing the curtains generally occurs between 21-24 o'clock because the user usually go to bed during this period. The dishwasher usually runs between 12-15 and 18-21 o'clock, which means that the user has lunch and dinner during this period, and then washes the dishes. As shown in Figure~\ref{fig:es}(b), users generally watch TV and do laundry on Saturdays and Sundays. As shown in Figure~\ref{fig:es}(c), users usually take a bath for about 1-2 hours, bath time longer than this may indicate abnormality occurs.

\section{Conclusion}

In this paper, we introduce \schemename \ for unsupervised user behavior anomaly detection. We first devise a Loss-guided Dynamic Mask Strategy (LDMS) to encourage the model to learn less frequent behaviors that are often overlooked during the learning process. Additionally, we introduce Three-level Time-aware Position Embedding (TTPE) to integrate temporal information into positional embedding, allowing for the detection of temporal context anomalies. Furthermore, we propose a Noise-aware Weighted Reconstruction Loss (NWRL) to assign distinct weights for routine behaviors and noise behaviors, thereby mitigating the impact of noise. Comprehensive experiments conducted on three datasets encompassing ten types of anomaly behaviors demonstrate that \schemename \ consistently outperforms state-of-the-art baselines while delivering highly interpretable results.

\begin{acks}
We thank the anonymous reviewers for their constructive feedback and comments. This work is supported by the Major Key Project of PCL under grant No. PCL2023A06-4, the National Key Research and Development Program of China under grant No. 2022YFB3105000, and the Shenzhen Key Lab
of Software Defined Networking under grant No. ZDSYS20140509172959989. \textbf{The first author, Jingyu Xiao, in particular, wants to thank
his parents Yingchao Xiao, Aiping Li and his girlfriend Liudi Shen for their kind support. He also thanks for all his friends who are the antidote during his tired period.}
\end{acks}





\balance

\bibliography{final}
\bibliographystyle{ACM-Reference-Format}

\newpage

\appendix

\section{Appendices}

\subsection{Notations}
Key notations used in the paper and their definitions are summarized in Table~\ref{tab:notation}.

\begin{table}[ht]
\caption{Main notations and their definitions.}
\label{tab:notation}
\setlength{\tabcolsep}{0.2em}{
\begin{tabular}{c|c}
\hline
Notation        & Definition                                                                                                                                          \\ \hline
$d$, $\mathcal{D}$ & a device/set of devices                                                                                                                 \\ \hline
$c$, $\mathcal{C}$        & a device control/set of device controls                                                 \\ \hline
$s$, $\mathcal{S}$  & a sequence/set of sequences                         \\ \hline
$n$  & the lenghth of a sequence                         \\ \hline
$b$         & a behavior                                                                                                                              \\ \hline
$t$ & the timestamp of behavior                                                                                                                                   \\ \hline
$hour$       & the hour of day of a behavior                                                                                                                             \\ \hline
$day$       & the day of week of a behavior                                              \\ \hline
$duration$        & the duration of a behavior \\ \hline
$PE$        & the positional embedding function \\ \hline
$PE(order)$, $w_{order}$        & the positional embedding and it's weight \\ \hline
$PE(hour)$, $w_{hour}$        & the hour embedding and it's weight \\ \hline
$PE(day)$, $w_{day}$        & the day embedding and it's weight \\ \hline
$PE(duration)$, $w_{dur}$        & the duration embedding and it's weight \\ \hline
$\overline{PE}$        & the positional embedding \\ \hline
$h_c$    & the device control embedding \\ \hline
$\mathbf{h}$    & the behavior embedding \\ \hline
$\mathcal{L}_{rec}$    & the reconstruction loss \\ \hline
$\ell_i$, $\mathcal{L}_{\text {vec}}$ & the loss of $i$-th behavior and loss vector\\ \hline
$w_i$, $\mathcal{W}_{\text {vec}}$ & the weight of $i$-th behavior and weight vector\\ \hline
$p_i$ & the normalized weight of $i$-th behavior\\ \hline
$mask$ & the mask vector\\ \hline
$score(s)$ & the anomaly score of sequence $s$\\ \hline
$th$ & the threshold\\ \hline
\end{tabular}}
\end{table}

\subsection{Device information of different dataset}
The AN, FR and SP data sets contain 36, 33, and 34 devices respectively, as shown in Table~\ref{tab:devInfo_an}, Table~\ref{tab:devInfo_fr}, and Table~\ref{tab:devInfo_sp}.

\begin{table}[ht]
\caption{Device information on AN dataset.}
\setlength{\tabcolsep}{.35em}{
\label{tab:devInfo_an} 
\begin{tabular}{cc|cc|cc}
\hline
No. & Device             & No. & Device         & No. & Device           \\ \hline
0   & AC                 & 12  & LED            & 24  & projector        \\
1   & heater             & 13  & locker         & 25  & washing\_machine \\
2   & dehumidifier       & 14  & bathheater     & 26  & kettle           \\
3   & humidifier\_1 & 15  & water\_cooler  & 27  & dishwasher       \\
4   & fan                & 16  & curtains       & 28  & bulb\_1     \\
5   & standheater        & 17  & outlet         & 29  & TV               \\
6   & aircleaner         & 18  & audio  & 30  & pet\_feeder      \\
7   & humidifier\_2  & 19  & plug     & 31  & hair\_dryer      \\
8   & desklight          & 20  & bulb\_2      & 32  & window\_cleaner  \\
9  & bedight\_1         & 21  & soundbox\_1 & 33  & bedlight\_2      \\
10  & camera             & 22  & soundbox\_2   & 34  & bedlight\_3      \\
11  & sweeper            & 23  & refrigerator   & 35  & cooler           \\ \hline
\end{tabular}}
\end{table}

\begin{table}[ht]
\caption{Device information on FR dataset.}
\setlength{\tabcolsep}{.10em}{
\label{tab:devInfo_fr} 
\begin{tabular}{cc|cc|cc}
\hline
No. & Device                  & No. & Device         & No. & Device           \\ \hline
0   & AirConditioner          & 11  & Fan            & 22  & Refrigerator        \\
1   & AirPurifier             & 12  & GarageDoor     & 23  & RemoteController \\
2   & Blind                   & 13  & Light          & 24  & RobotCleaner           \\
3   & Camera                  & 14  & Microwave      & 25  & Siren      \\
4   & ClothingCareMachine     & 15  & MotionSensor   & 26  & SmartLock     \\
5   & Computer                & 16  & NetworkAudio   & 27  & SmartPlug               \\
6   & ContactSensor           & 17  & None           & 28  & Switch      \\
7   & CurbPowerMeter          & 18  & Other          & 29  & Television      \\
8   & Dishwasher              & 19  & Oven           & 30  & Thermostat  \\
9  & Dryer                    & 20  & PresenceSensor & 31  & Washer      \\
10  & Elevator                & 21  & Projector      & 32  & WaterValve      \\ \hline
\end{tabular}}
\end{table}

\begin{table}[ht]
\caption{Device information on SP dataset.}
\setlength{\tabcolsep}{.10em}{
\label{tab:devInfo_sp} 
\begin{tabular}{cc|cc|cc}
\hline
No. & Device                  & No. & Device         & No. & Device           \\ \hline
0   & AirConditioner          & 12  & GarageDoor     & 24  & RobotCleaner         \\
1   & AirPurifier             & 13  & Light          & 25  & SetTop \\
2   & Blind                   & 14  & Microwave      & 26  & Siren            \\
3   & Camera                  & 15  & MotionSensor   & 27  & SmartLock     \\
4   & ClothingCareMachine     & 16  & NetworkAudio   & 28  & SmartPlug    \\
5   & Computer                & 17  & None           & 29  & Switch               \\
6   & ContactSensor           & 18  & Other          & 30  & Television     \\
7   & CurbPowerMeter          & 19  & Oven           & 31  & Thermostat      \\
8   & Dishwasher              & 20  & PresenceSensor & 32  & Washer  \\
9  & Dryer                    & 21  & Projector      & 33  & WaterValve      \\
10  & Elevator                & 22  & Refrigerator                        \\ 
11  & Fan                     & 23  & RemoteController                    \\ \hline
\end{tabular}}
\end{table}

\subsection{Data collection}
\begin{figure}[ht]
\centering
\includegraphics[width = .45\textwidth]{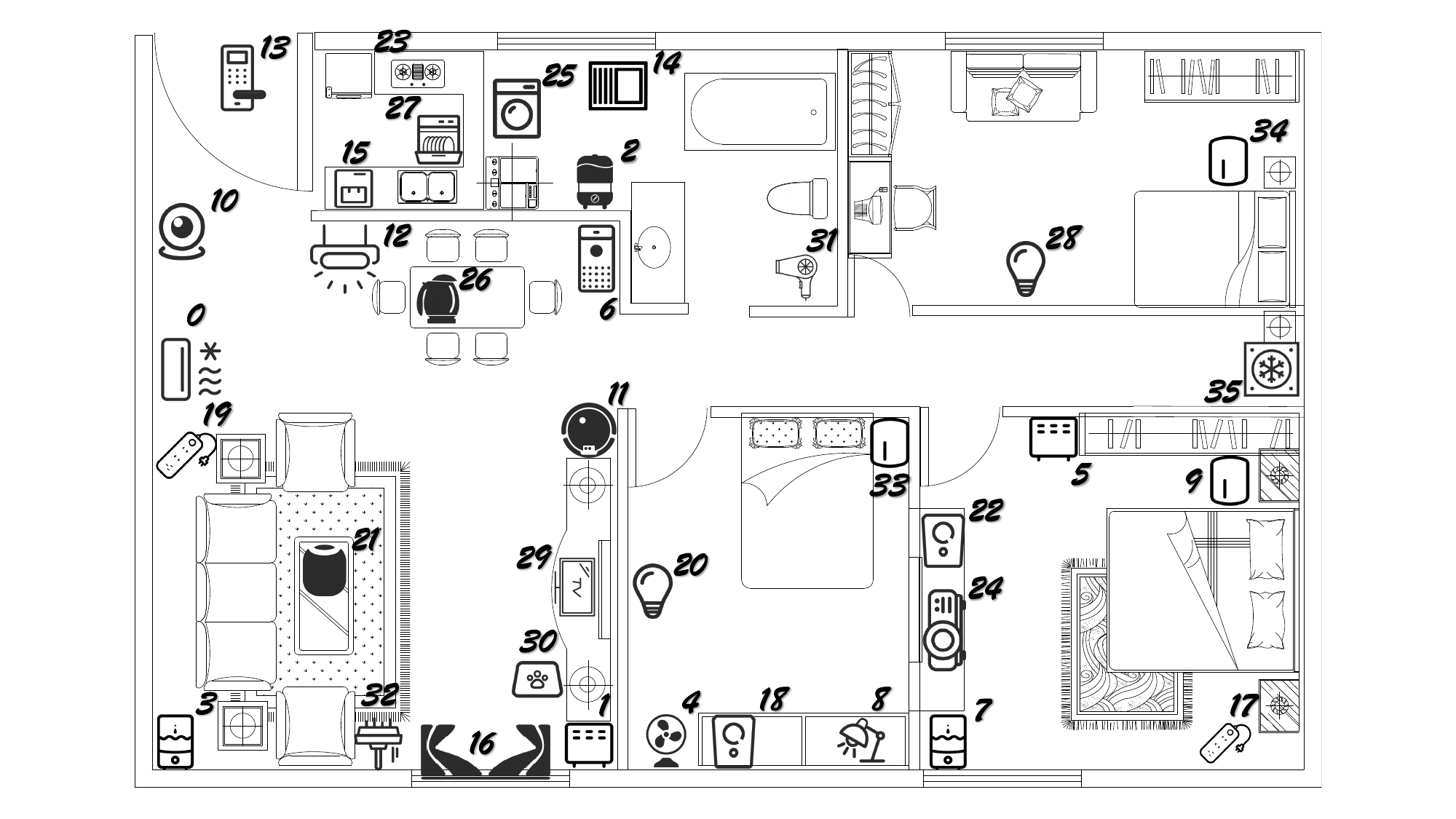}
\caption{Overview of testbed setup.}
\label{fig:floorplan}
\end{figure}

\textbf{Testbed and Participants.} To create a practical and viable smart home model, we implemented our experimental platform within an apartment setting to gather user usage data of various devices, forming our smart home user behavior dataset (AN). Three volunteers were recruited to simulate the typical daily activities of a standard family, assuming the roles of an adult male, an adult female, and a child. The experimental platform comprises a comprehensive selection of 36 popular market-available devices, detailed in Table~\ref{tab:devInfo_an}, with their deployment illustrated in Figure~\ref{fig:floorplan}.


\textbf{Normal Behavior Collection.} We enlisted volunteers to reside in apartments and encouraged them to utilize equipment in accordance with their individual habits. Throughout the designated period of occupancy, we refrained from actively or directly intervening in the users' behavior. However, we implemented a system where users consistently logged their activities. Following the conclusion of the data collection phase, we reviewed the device usage logs via the smart home app, amalgamating these logs with the users' behavior records to compile a comprehensive user behavior dataset. To mitigate potential biases arising from acclimating to a new living environment, participants were required to inhabit the experimental setting for a minimum of two weeks before the formal commencement of data collection. All users possessed comprehensive knowledge of the IoT devices and applications in use. Subsequent to check-in, control of all devices was relinquished to the users, who were duly informed in advance that their device usage would be subsequently reviewed and analyzed by our team

\begin{figure}[ht]
\centering
\includegraphics[width = .42\textwidth]{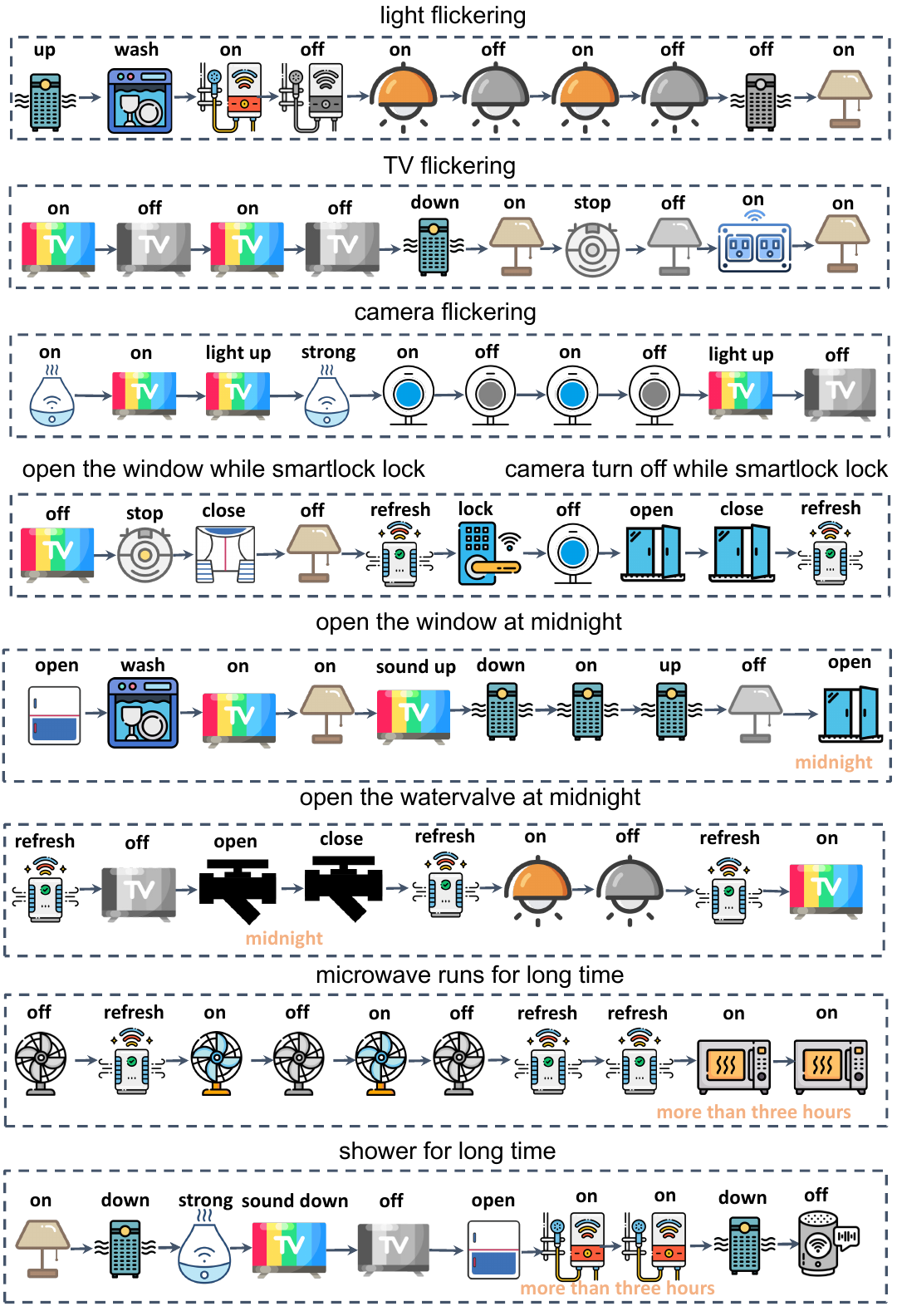}
\caption{Examples of anomaly behavior sequences.}
\label{fig:anomaly_example}
\end{figure}

\textbf{Anomaly Behavior Injection.} We insert abnormal behaviors in Table~\ref{tab:types} into normal behavior sequences to construct abnormal behavior sequences. Then the abnormal behavior sequences. Then the abnormal behavior sequence and the normal behavior sequence together form the test dataset. The anomaly behavior sequences examples are shown in Figure~\ref{fig:anomaly_example}.

\subsection{Detailed experimental settings}

All models (including baselines and \schemename) are implemented by PyTorch \cite{pytorch} and run on a graphic card of GeForce RTX 3090 Ti. All models are trained with Adam optimizer \cite{kingma2014adam} with learning rate 0.001. We train \schemename \ to minimize $\mathcal{L}_{rec}$ in Equation (\ref{equ:obj}). During training, we monitor reconstruction loss and stop training if there is no performance improvement on the validation set in 10 steps. For model hyperparameters of \schemename, we set the batch size to 512 and the initial weights of TTPE are $w_{order}=0.1, w_{hour}=0.4, w_{day}=0.4$, and $w_{duration}=0.7$. For mask ratio and step without step, we search in $\{0.2, 0.4, 0.6, 0.8\}$ and $\{3, 4, 5, 6\}$, respectively. We chose the number of encoder and decoder layers in $\{1, 2, 3, 4\}$, and the embedding size in $\{8, 16, 32, 64, 128, 256, 512\}$.

\subsection{Mask strategy deep dive} 

\begin{figure}[ht]
    \subfigure[F1-Score.]{
    \label{fig:mask_per}
    \centering
    \includegraphics[width = .22\textwidth]{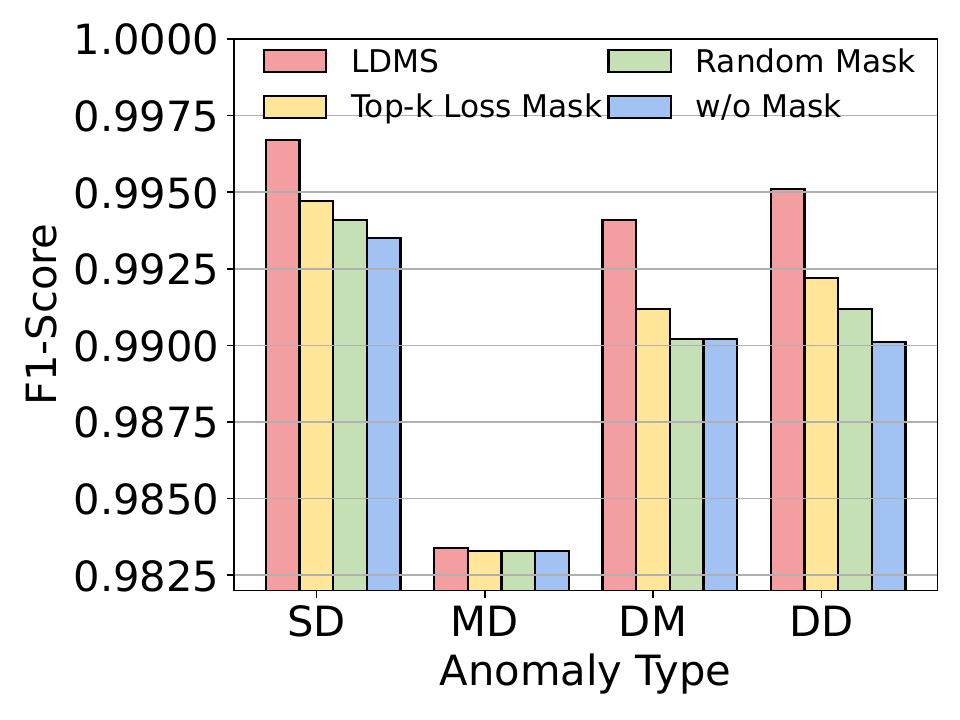}
    }
    \subfigure[Variance.]{
    \label{fig:mask_var}
    \centering
    \includegraphics[width = .22\textwidth]{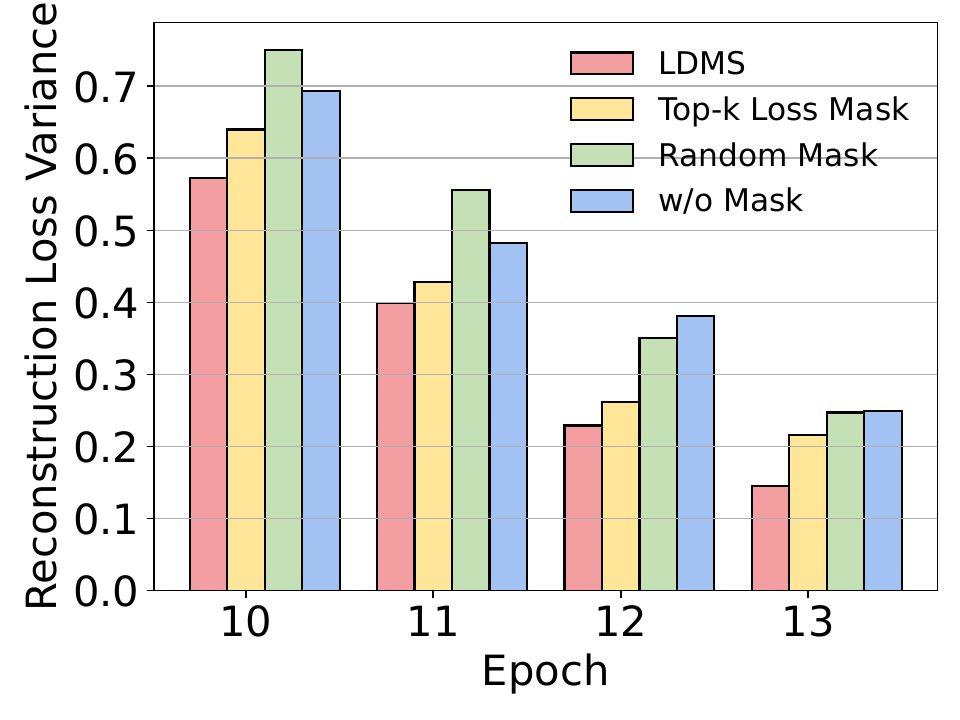}
    }
    \caption{Performance and reconstruction loss variance on AN dataset under different mask strategy.}
\end{figure}

\begin{figure}[ht]
\centering
\includegraphics[width = .4\textwidth]{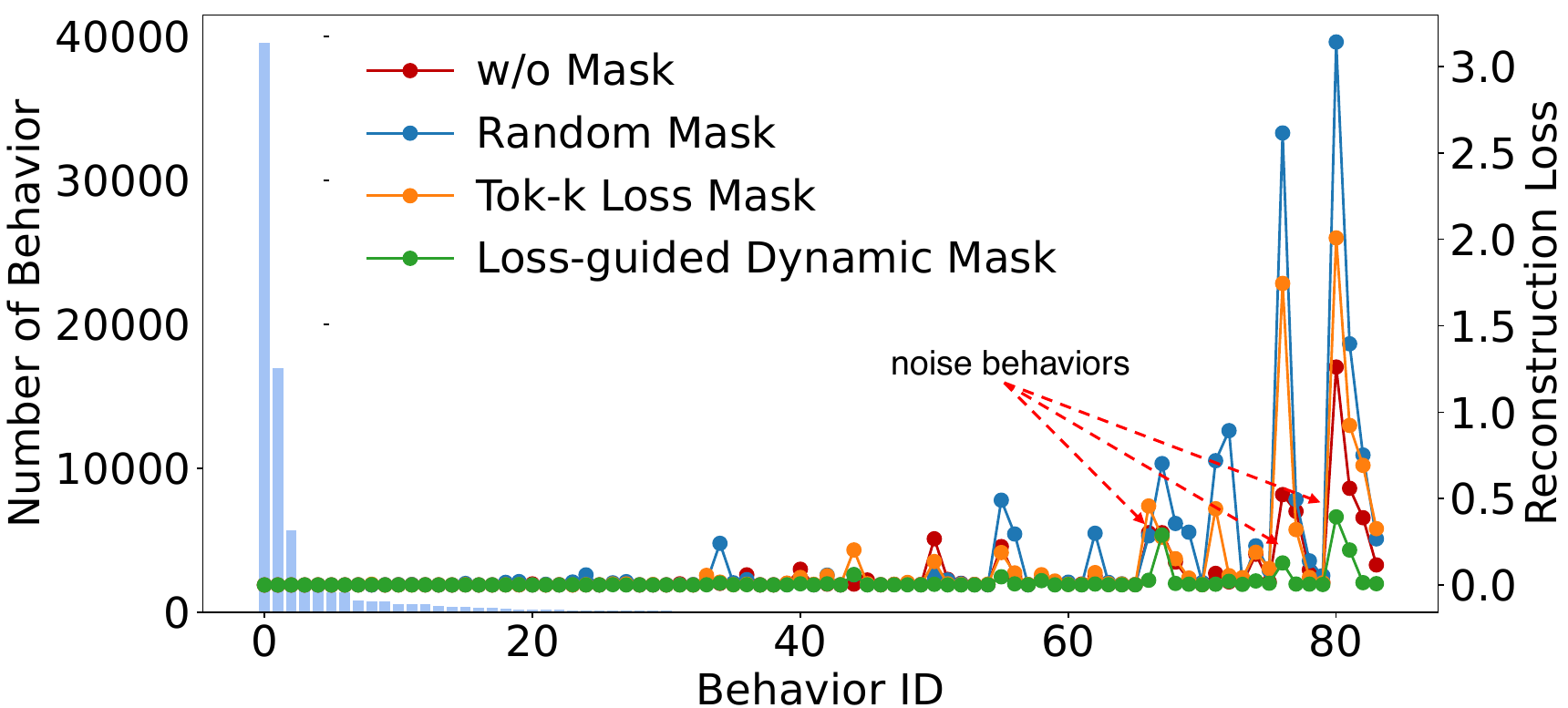}
\caption{Reconstruction loss distribution under different mask strategy on SP dataset.}
\label{fig:mask_dis}
\end{figure}

To verify the effectiveness of LDMS, we compared it with the three baselines (w/o mask, random mask and top-$k$ loss mask)  mentioned in the section \ref{subsec:LDMS}. As illustrated in Figure~\ref{fig:mask_per}, LDMS consistently outperforms all other mask strategies across four types of anomalies. The results presented in Figure~\ref{fig:mask_var} further demonstrate that LDMS exhibits the smallest variance in reconstruction loss throughout the training process, which demonstrates that \schemename \ learns both \textit{easy-to-learn} behaviors and \textit{hard-to-learn} behaviors very well. We also plotted the loss distribution diagram under different mask strategies. As shown in Figure~\ref{fig:mask_dis}, LDMS shows the smallest reconstruction loss and variance, which demonstrates that our mask strategy can better learn \textit{hard-to-learn} behaviors. We can still observe behaviors with high reconstruction loss as pointed by the red dashed arrow after applying LDMS, which is likely to be noise behaviors, thus it's necessary to assign small weights for these noise behaviors during anomaly detection for avoiding identifying normal sequences containing noise behaviors as abnormal.



\end{document}